\pdfoutput=1
\documentclass{article}

\usepackage{amsmath,amssymb,longtable,hyperref,bbm}

\usepackage[pdftex]{graphicx}
\usepackage[pdftex]{color}

\def\be{\begin{equation}}
\def\beq{\begin{equation}}
\def\ee{\end{equation}}
\def\eeq{\end{equation}}
\def\ba{\begin{eqnarray}}
\def\ea{\end{eqnarray}}

\newcommand{\E}{{\cal E}}

\newcommand{\eg}{{\it e.g.~}}
\newcommand{\ie}{{\it i.e.~}}

\def\one{\mathbbm{1}}

\def\IP{\relax{\rm I\kern-.18em P}}

\def\eqn#1{\begin{equation}\begin{split}#1\end{split}\end{equation}}
\def\eqna#1{\begin{eqnarray}#1\end{eqnarray}}
\def\<{\langle}
\def\>{\rangle}
\def\vev#1{\left\langle #1 \right\rangle}

\begin{document}

\title{Moduli Webs and Superpotentials for Five-Branes}  
\author{
\large
{\bf Marco Baumgartl}\footnote{M.Baumgartl@physik.uni-muenchen.de, marco.baumgartl@universe-cluster.de}\\
\it
\normalsize\it Arnold-Sommerfeld Center for Theoretical Physics,\\
\normalsize\it LMU Munich, Theresienstr. 37, D--80333 Munich\\
\normalsize\rm and \it The Cluster of Excellence for Fundamental Physics,\\
\normalsize\it Boltzmannstr. 2, D--85748 Garching\\[12pt]
\large
{\bf Simon Wood}\footnote{swood@phys.ethz.ch}\\
\normalsize\it Institut f{\"u}r Theoretische Physik, ETH Z\"urich,\\
\normalsize\it CH--8093 Z{\"u}rich, Switzerland
}

\date{}

\maketitle

\abstract{We investigate the one-parameter Calabi-Yau models and identify families of D5-branes which are associated to lines embedded in these manifolds. The moduli spaces are given by sets of Riemann curves, which form a web whose intersection points are described by permutation branes. We arrive at a geometric interpretation for bulk-boundary correlators as holomorphic differentials on the moduli space and use this to compute effective open-closed superpotentials to all orders in the open string couplings. The fixed points of D5-brane moduli under bulk deformations are determined.} 
\vskip 11pt
\noindent
LMU-ASC-63/08

\newpage
\tableofcontents

\newpage

\section{Introduction}

The matter content of string theory depends on the compactification details of higher dimensions as well as the brane configurations in the hidden dimensions. Often such configurations are organised in families, which constitute some open string moduli space. 
In order to better understand the role of the moduli space it is important to go beyond perturbative calculations and examine its global geometry. 
In this paper we will study the moduli spaces of D5-branes in ten-dimensional string theory compactified on various specific Calabi-Yau manifolds, which are constructed of tensor products of minimal models. 
We compute exact superpotentials for open string moduli under bulk perturbations and find explicit conditions for branes in order to survive the large complex structure limit.
The moduli space turns out to be a web of intersecting complex lines with a generic structure, whose intersections are permutation points.
We construct this space exactly to all orders in perturbation theory, investigate its symmetry properties, compute the marginal spectrum and find joining relations for the transitions of cohomologies at the intersection points.

We work in the B-model in a stringy regime at the Landau-Ginzburg point, where
the K\"ahler moduli are decoupled. At this point the model is realised as an (orbifold of) a Landau-Ginzburg theory.  
Our results are achieved by matrix factorisation technologies.
Matrix factorisations are establishing a novel way in the study of open strings on Calabi-Yau spaces. This technology can be applied at the Landau-Ginzburg point, where boundary degrees of freedom have a matrix representation. Concretely we will work with the topologically B-twisted model, which restricts the theory to its BPS spectrum. Of particular interest will be the boundary part $Q$ of the BRST operator. This operator is related to (superpositions of) D-branes, whose connection to the conformal field theory description of minimal models and Gepner models is well understood. D5-branes with three extended directions and two directions in the compactified space (which appear as two-dimensional branes in the Calabi-Yau) have a geometric meaning as complex lines in the Calabi-Yau manifold. The deformation space of such lines is identical to the moduli space of these D5-branes.

Conformal field theory descriptions are usually only available for certain points in moduli space. Moving away from such points is technically difficult and involves the construction of renormalisation group flows. Matrix factorisations provide an alternative description, where one has access to the chiral primary fields in the spectrum only.
One of their attractive features lies in the fact that in some cases it is
possible to study deformations which allow the exploration of connected
regions in moduli space. A good understanding of moduli spaces and the
relation between open and closed string moduli is essential for acquiring
insight into the structure of full string theory. The coupled open-closed
superpotential for D-branes on Calabi-Yau manifolds is much sought after. It is interesting from a phenomenological point of view since it determines various string couplings. It also plays an important role in approaches to open mirror symmetry, including open-closed Picard-Fuchs equations and relative period integrals \cite{Aganagic:2000gs,Aganagic:2001nx,Brunner:2006tc,Kachru:2000an,Klemm:1992tx,Knapp:2008uw,Lerche:2001cw,Lerche:2002yw,Mayr:2001xk,Walcher:2006rs}. Beyond this it enters the discussion of background independence of string theory along the lines of \cite{Baumgartl:2004iy, Baumgartl:2006xb, Herbst:2004zm,Neitzke:2007yw}.

Occasions are rare where one is actually able to cover not only infinitesimal parts of moduli spaces. In this paper we will use geometrical methods in order to find matrix factorisation descriptions for D-branes at the Gepner point. Such factorisations are then constructed over the whole open string moduli space to all orders in the boundary couplings. Following the methods developed in \cite{Baumgartl:2007an}  we explicitely compute these spaces for the quintic and the one-parameter family $\mathbb P_{(1,1,1,1,2)}[6]$, $\mathbb P_{(1,1,1,1,4)}[8]$ and $\mathbb P_{(1,1,1,2,5)}[10]$. 

Under bulk deformations the boundary moduli space can change significantly
(see for example \cite{Fredenhagen:2006dn}). We show that bulk-boundary correlators are in correspondence to holomorphic differentials on the moduli space, which is a result important for integrability of three-point functions and therefore for the existence of effective potentials. We are able to identify those points in moduli space for which matrix factorisations deform with complex structure deformations, thus representing marginal directions in the open-closed moduli space. Under such deformations the boundary moduli space collapses to a discrete set of points, fixing the open string moduli as functions of the closed string moduli. In addition correlators between boundary fields and marginal bulk fields contain information about the effective superpotential.
Since we know this correlator at any point in the boundary moduli space, this allows us to integrate it and obtain an expression for the effective superpotential. This result is exact in open string couplings but first order in closed string couplings.
Effective superpotentials have been perturbatively computed in \cite{Ashok:2004xq,Brunner:1999jq,Brunner:2000wx,Douglas:2002fr,Govindarajan:2006uy,Herbst:2004ax,Herbst:2004zm,HLL,Kachru:2000ih}

We start with a brief summary of methods and results of previous work in the next section, where matrix factorisations are introduced and their connections to BCFT boundary states are tersely outlined. Section \ref{CY0} is devoted to the Fermat quintic. Techniques which are important in later sections are introduced here. The marginal cohomologies are computed. We look at the symmetries of intersections and show how the moduli web emerges from joining relations between the moduli branches. The bulk-induced superpotential is computed. In section \ref{sec:P6} we focus on the threefold $\mathbb P_{(1,1,1,1,2)}[6]$. Due to the different weights of the coordinates, which is reflected in the spectrum, the joining relations are modified while the global structure of the moduli web is unchanged. We obtain again expressions for the effective superpotential and verify the correspondence between bulk deformations and holomorphic differentials on the moduli space. Very similar results are obtained in section \ref{CY2} while here for the first time isolated marginal states are observed which only live in an enhanced spectrum at some permutation points.
In section \ref{CY3} we discuss the more intricate case of the $\mathbb P_{(1,1,1,2,5)}[10]$ threefold. The different weights in this model introduce more complexity. Cohomologies, joining relations and superpotentials are determined as in the previous models.

\subsection{Matrix factorisations}
\label{MFintro}

Before we start with the quintic we repeat the basic construction of matrix factorisations for minimal models, their tensor products and (generalised) permutation branes.

Matrix factorisations arise in the context of $N=2$ Landau-Ginzburg models with superpotential $W$ and are important to understand the connection of string theory on Calabi-Yau manifolds with minimal models and Gepner models \cite{DVV, Gepner:1986wi,Gepner:1987qi,Gepner:1987vz,Greene:1988ut,  Lerche:1989uy,Witten:1993yc, HHP} in the presence of D-branes. The presence of a worldsheet boundary breaks $N=2$ supersymmetry so that only one supersymmetry charge is preserved. There are two distinct ways to combine the two left and right moving bulk supercharges into a boundary supercharge, resulting in A- or B-type supersymmetry. We will focus on the latter.

Supersymmetric boundary conditions for open strings can be obtained along the route described in \cite{Warner:1995ay} (see also \cite{Hori:2000ic}). This approach utilises the fact that on an open string worldsheet it is possible to introduce fermionic boundary fields $\pi, \bar\pi$ \cite{Lazaroiu:2003zi,Witten:1998cd}. It is in fact necessary to include fermionic boundary terms in order to cancel boundary contributions to supersymmetry variations in the bulk. The boundary fermions together with the fields coming from the bulk are the building blocks of the open string BPS spectrum. This spectrum is obtained as the cohomology of a boundary part $Q$ of the supersymmetry charge. $Q$ can be expressed as\footnote{In general higher powers of $\pi$ and $\bar\pi$ can appear. In this article it will only be necessary to consider operators which are products of $Q$s linear in the boundary fermions.}
\eqn{
	Q=\sum_i \left(\pi_iJ_i + \bar\pi_iE_i\right)\ ,
}
where $J$ and $E$ are polynomials of the bosonic fields\footnote{We will eventually use the notation $(J_1, E_1)$ for $\pi_1J_1 + \bar\pi_1E_1$. Also we will denote graded tensor products as $(J_1, E_1)\odot(J_2,E_2) = \sum_{i=1}^2 \left(\pi_iJ_i + \bar\pi_iE_i\right)$, where a suitable choice of the matrix representation of $\pi^i$ and $\bar\pi^i$ is understood.}. The supersymmetry condition becomes
\eqn{\label{eqnMF}
	Q^2 = W
}
or
\eqn{
	W=\sum_i J_iE_i\ .
}
Since $\pi$ and $\bar\pi$ have a Clifford representation as graded matrices $Q$ can also be represented as a matrix. Equation (\ref{eqnMF}) can then be viewed as a matrix equation in which $Q$ is the square root of the superpotential \cite{Ashok:2004zb, Aspinwall:2006ib, Aspinwall:2007cs,Brunner:2003dc,Douglas:2002fr,Kapustin:2002bi,Kapustin:2003rc}.

\subsection{Three-point functions and bulk-boundary correlators}
\label{sec:correl}

A central tool for our calculations is the Kapustin-Li formula derived
in \cite{Kapustin:2003ga,Brunner:2003dc}. It
allows one to calculate three-point functions and bulk-boundary
correlators. For a bulk field $\Phi$ and a boundary field $\psi$ the
formula is
\eqn{
  \langle \Phi\psi\rangle=\text{Res}\Phi\frac{\text{STr}[\partial_{x_1}Q\cdots\partial_{x_5}Q\psi]}{\partial_{x_1}W\cdots\partial_{x_5}W}\ .
}
For three boundary fields $\psi_1,\psi_2,\psi_3$ it is
\eqn{
  \langle \psi_1\psi_2\psi_3\rangle=\text{Res}\frac{\text{STr}[\partial_{x_1}Q\cdots\partial_{x_5}Q\psi_1\psi_2\psi_3]}{\partial_{x_1}W\cdots\partial_{x_5}W}\ ,
}
where the residue is taken at the critical points of $W$. We will use this formula later in order to determine bulk-boundary couplings as derivatives of an open-closed superpotential.

\subsection{Minimal models}

The simplest models which allow non-trivial matrix factorisations are the minimal models of type $A_{d-2}$. These are related to Landau-Ginzburg models with $W=x^{d}$. The spectrum of D-branes obtained through matrix factorisations is given by a set $Q_n=\pi x^n + \bar\pi x^{d-n}$ where $n\le\left[\frac{d}{2}\right]$. Choosing a matrix representation together with a grading operator $\sigma=\text{diag}(-1,1)$ gives a family
\eqn{
	Q_n=\begin{pmatrix}0&J\\E&0\end{pmatrix}=\begin{pmatrix}0&x^n\\x^{d-n}&0\end{pmatrix}\ .
}
In the conformal field theory language, $Q_n$ corresponds to the boundary state $|L,S\,\rangle\!\rangle = |n-1,0\,\rangle\!\rangle$ in the B-model \cite{Brunner:2003dc,Kapustin:2003rc}.
The BPS spectrum of strings $\Psi$ between two D-branes with $Q$ and $Q'$ is then obtained as cohomology of the twisted differential
\eqn{
	D\Psi=Q\Psi - (-1)^{|\Psi|}\Psi Q'\ ,
}
where $|\Psi|$ is the fermion number of the field $\Psi$. We restrict ourselves to the spectrum of a single D-brane, so $Q'=Q$. For the $A_k$ minimal model with only one such D-brane the fermions are given by
\eqn{
	\Psi_l=\begin{pmatrix}0&x^{l}\\-x^{d-2n+l}&0\end{pmatrix}
}
and the bosons by
\eqn{
	\Phi_l=\begin{pmatrix}x^l&0\\0&x^l\end{pmatrix}
}
with $0\le l<n<d$. Thus there are $n$ fermions and bosons in the spectrum. 

It is helpful to keep track of the R-charges of the various states. In the bulk the superpotential is normalised to charge 2, which fixes the R-charges of the chiral bulk fields to $[x]=\frac{2}{d}$. At the boundary $Q$ must have charge 1, due to (\ref{eqnMF}). From this it is easy to write down the charges for the boundary fermions to 
$[\pi]=1-\frac{2n}{d} J=-[\bar\pi]$, where $n=$deg $J$ is the degree of the homogenous polynomial $J$. We will focus on deg$J=1$ throughout this paper, in which case $[\pi]=\frac{d-2}{d}$.
Therefore $[\Phi_l] = \frac{2l}{d}$ and $[\Psi_l]=\frac{d-2n+2l}{d}$.

This construction can be extended to more complicated models. 
For tensor products of minimal models higher-dimensional representations of the Clifford-algebra must be used\footnote{For the description of Gepner models we must in addition orbifold, but this will be of no relevance for our further computations.}. The BRST operator of the tensored theories becomes a graded tensor product of the BRST operators associated to each of the building blocks. 

To describe Calabi-Yau compactifications at the stringy point, one must consider orbifolds of graded tensor products. The orbifold projects on integer charges in the bulk and is necessary to conduct a GSO projection of the theory. For D-branes, orbifolding introduces an extra representation label that has been discussed in \cite{Ashok:2004zb} in the context of Landau-Ginzburg models. In this paper we will consider only single branes; in this case the projection is on integer charges, also in the boundary sector.

\subsection{Permutation branes}

In the following, permutation branes and generalised permutation branes are of some importance, thus we will summarise some facts about them.

Permutation branes have been constructed as objects in CFT in \cite{Recknagel:1998ih, Recknagel:2002qq}. Their matrix factorisation representation has been described \eg in \cite{Brunner:2004mt, Brunner:2005fv, Enger:2005jk, Govindarajan:2005im}. They correspond to conformal boundary conditions which exchange the currents of tensored minimal models at the boundary. They are of the form\footnote{When $x_1$ and $x_2$ do not appear with the same exponent, we call them generalised permutation branes.}  \cite{Brunner:2005fv}
\eqn{
	J_{ML} = \prod_{m=(M-L)/2}^{(M+L)/2}(x_1-\eta_m x_2)
}
with $\eta_m=e^{-\pi i\frac{2m+1}{d}}$ a $d$-th root of $-1$ and $W=x_1^d+x_2^d$.
In conformal field theory language this translates into the boundary state
\eqn{
	J_{ML} \qquad\Leftrightarrow\qquad |L,M,S_1=0,S_2=0\,\rangle\!\rangle\ .
}
As we will be interested in linear matrix factorisations, $J$ will always be a polynomial of degree 1, so $J=x_1-\eta x_2$, where $\eta$ stands for one of the roots $\eta_m$. 

For the computation of the spectrum we will consider generalised permutation branes. Let $W$ be of the form
\eqn{W(x_1,x_2)=x_1^{dn}+x_2^{dm}\ ,}
where $n$ and $m$ are coprime. The linear factorisations are then given
\eqn{
	J=x_1^n - \eta x_2^m\qquad E=\prod_{\eta'\ne\eta}\left(x_1^n - \eta x_2^m\right)\ ,
}
where $\eta^d=\eta'^d=-1$. The cohomology is easily calculated. It contains no fermions. The bosons are of the form
\eqn{
	\Phi_{ij}=x_1^ix_2^j\one\ ,
}
with the constraints $0\le i<n(d-1)-1$ and $0\le j<m$. Thus there are $(nd-n-1)m$ of them and their charges are $[\Phi_{ij}]=\frac{2i}{nd}+\frac{2j}{md}$.

In the following we will use tensor products of minimal modes boundary states and permutation boundary states in order to construct the D5-branes in various Calabi-Yau spaces.
In particular we will use them to compute superpotentials on these manifolds.

\subsection{Equivalence classes}

Matrix factorisations are equipped with an obvious gauge freedom. The supersymmetry condition $Q^2=W\cdot\one$ is invariant under similarity transformations $Q \sim U Q U^{-1}$. The matrix $U$ has even grading and appears therefore in block diagonal form
\eqn{
	U = U_0\oplus U_1 = \begin{pmatrix}U_0& 0\\0&U_1\end{pmatrix}\ .
}
The matrices $U_0$ and $U_1$ must be invertible over the polynomial ring. In particular, all constant invertible matrices represent gauge transformations which also contain standard row and column operations \cite{Govindarajan:2005im}.

Gauge transformations affect the form of $Q$ as well as the expressions for the cohomology elements, but physical data of the brane are unaffected.

A second equivalence relation is given by adding and subtracting trivial factorisations. The factorisations $W=1\cdot W$ is trivial in the sense that its cohomology is empty. It indeed describes rather the situation when a boundary term in the Landau-Ginzburg action is absent (or trivially decoupled) and must therefore be identified with the braneless vacuum. Physical data of a brane are independent or the operation \cite{Herbst:2004zm}
\eqn{
	Q\sim Q\oplus\begin{pmatrix}0&1\\W&0\end{pmatrix}\ .
} 

In the following we will sometimes fix the gauge and choose a particular representative of the equivalence class.

\section{The Fermat Quintic}
\label{CY0}
\subsection{D-branes families}

We begin by considering D-branes in the Fermat quintic, which is given as the geometrical zero locus of the polynomial
\eqn{W=x_1^5+x_2^5+x_3^5+x_4^5+x_5^5\ .}
This model and its brane superpotentials have been studied before many times, for example in \cite{Ashok:2004xq,Brunner:1999jq}. Many branes of this model are known and their connection to boundary states in the corresponding Gepner model have been worked out \cite{Ashok:2004xq, Ashok:2004zb, Hori:2004ja, Brunner:2003dc}. It has been shown, basically by counting intersections \cite{Brunner:2005fv} and also by more general arguments \cite{O,HHP}, that linear permutation branes in this model correspond to D5-branes. Geometrically these are complex lines in the projective space where $W=0$.

There is a straightforward way to associate a matrix factorisation to a given line in the Calabi-Yau manifold. As an example consider the line given by the intersection of the three polynomials $J_1=x_1-\eta x_2$, $J_2=x_3-\eta'x_4$, $J_3=x_5$, where $\eta^5=\eta'^5=-1$. Employing the {\em Nullstellensatz} we know that $W$ can be factored as \cite{Brunner:2005fv,HHP,O}
\eqn{W=J_1E_1+J_2E_2+J_3E_3\ ,}
and it is easy to write down such polynomials $E_i$ by using $x^5+x'^5=\prod_{i=1}^5(x-\eta_ix')$. Since these pairs $(J_i,E_i)$ are exactly the data we need to construct a matrix factorisation, we can immediately write down the result:
\eqn{
	Q=Q_1\odot Q_2\odot Q_3=\sum_{i=1}^3\left(J_i\pi_i+E_i\bar\pi_i\right)\ .
}

Let us denote the matrix factorisation $Q_1=J_1\pi_1+E_1\bar\pi_1$ by $(12)$, since it mixes the first and the second coordinate. Then a short hand notation for the above brane is $(12)(34)(5)$. We can use the symmetry group of the quintic, which is just the symmetric group $S_5$, to permute the coordinates and construct more such branes. This way all the permutation D5-branes can be generated.

It has been shown in \cite{AK} that there are many such complex lines and that they are organised in one-parameter families. They also give rise to branes wrapping these lines, which have been discussed in \cite{Baumgartl:2007an}. Under generic bulk deformations not all lines in a family will adjust to the new complex structure. Only a small set which does not break supersymmetry will be left. We will interpret this later as a collapse of the moduli space through bulk-induced
lifting of boundary moduli.

We start with the ansatz	
\eqn{\label{quinticlines} l = (u: \eta u: va: vb: vc)\ .}
Here $(u: v)\in \mathbb{P}^1$.  The parameters $(a:b:c)\in\mathbb{P}^2$ must be chosen in a way so that the line $l$ lies in $W$. The condition we obtain from this is
\eqn{\label{eqnQuinticModuli}a^5+b^5+c^5=0\ .}
This is the Riemann surface describing the moduli space of the lines of the form (\ref{quinticlines}). 
Note that since $\eta$ is a 5th root of $-1$, there are five copies of each of these Riemann surfaces.

Now we can use (\ref{quinticlines}) to read off the corresponding matrix factorisations: 
\eqn{
\label{qMF}
	J_1=x_1-\eta x_2\qquad J_2=ax_4-bx_3\qquad J_3=cx_3-ax_5\ .}
The associated polynomials $E_i$ are quoted in appendix \ref{App1Fac}.

Inserting (\ref{qMF}) into (\ref{eqnMF}) 
yields the same condition (\ref{eqnQuinticModuli}) on the moduli as in the purely geometric treatment. Since $Q=Q(a,b,c)$ depends parametrically on the moduli, we have constructed a family of BRST-operators defined smoothly over the whole moduli space.  
Of course there are many such matrix factorisations, such as 
\eqn{
\label{qMF2}
	J_1=x_1-\eta x_2\qquad J_2=ax_5-cx_3\qquad J_3=cx_4-bx_5\ ,
}
which are associated to the same D5-brane. It can be convenient to switch between representations of the BRST operator when approaching other patches of the moduli space. For example, (\ref{qMF}) is a good factorisation in the patch $a=1$ and we can evaluate $Q$ at the points $b=0$ and $c=0$. In order to describe the patch $b=1$ the factorisation (\ref{qMF2}) is more suitable. Both sets of polynomials describe the same physical quantity, so they are representatives of the same equivalence class of matrix factorisations related by gauge transformations. 

In the following it will sometimes be necessary to explicitly choose a particular gauge. To keep track of the combinatorics we find it helpful to introduce the following
\paragraph{Notation:}
The expression $(i)$ denotes the linear matrix factorisation for a minimal model in the $i$th coordinate.
The expression $(ij)$ denotes the linear matrix factorisation $J = x_i-\eta x_j$, with appropriate $\eta$ (in particular, $(ij)$ and $(ji)$ are gauge equivalent but not identical).
The expression $(ijk)$ denotes the linear matrix factorisation defined by $J=ax_i-bx_j$ and $J'=cx_j-ax_k$. Note that this implicitly fixes how $a$, $b$ and $c$ appear in the polynomials. All permutations $(\sigma(i)\sigma(j)\sigma(k))$ describe gauge equivalent matrix factorisations. We will not further distinguish between matrix factorisations $(ij)(klm)$ and $(klm)(ij)$ etc.
\\

The factorisation (\ref{qMF}) is thus given and fixed by the expression $(12)(435)$. This notation encodes the symmetries of (\ref{quinticlines}) nicely. It is easy to see, that there is also a matrix factorisation $(34)(215)$, for instance, which corresponds to
\eqn{
	J'_1=x_3-\eta' x_4\qquad J'_2=a'x_2-b'x_1\qquad J'_3=c'x_1-a'x_5\ .
}
In addition it is easy to find the intersection of branches in the moduli space. For example the branch $(12)(435)$ contains the three special points $(12)(43)(5)$, $(12)(35)(4)$ and $(12)(45)(3)$. As we have seen above these are permutation points. The point $(12)(43)(5)$ is also part of the branch $(34)(215)$ hence it is an intersection point between the branches (given that the four roots of $-1$ are chosen correctly). This is enough to set up a list of all branches and their intersections:

\begin{table}[htdp]
\centering
\begin{tabular}{|c|c|c|}
\hline
name & factorisation & intersects with \\
\hline
$(\alpha)	$&$(12)(435)$&$(\beta), (\zeta), (\rho)$\\
$(\beta)		$&$(35)(412)$&$(\alpha), (\gamma), (\mu)$\\
$(\gamma)	$&$(14)(325)$&$(\beta), (\delta), (\nu)$\\
$(\delta)	$&$(23)(415)$&$(\gamma), (\epsilon), (\rho)$\\
$(\epsilon)	$&$(15)(324)$&$(\delta), (\zeta), (\mu)$\\
$(\zeta)		$&$(34)(215)$&$(\epsilon), (\alpha), (\nu)$\\
$(\lambda)	$&$(13)(245)$&$(\mu), (\nu), (\rho)$\\
$(\mu)		$&$(24)(315)$&$(\beta), (\lambda), (\epsilon)$\\
$(\nu)		$&$(25)(134)$&$(\gamma), (\zeta), (\lambda)$\\
$(\rho)		$&$(45)(123)$&$(\alpha), (\delta)	, (\lambda)$\\
\hline
\end{tabular}
\caption{Complete list of branches and their intersections}
\label{branches}
\end{table}
These are all possible ${5 \choose 2}=10$ families of lines. Note that permutation points are those points on moduli space which lie on exactly two branches, so we can identify them by specifying the two intersecting branches. Thus we will refer to them for example as $P_{(\alpha\beta)}=(12)(43)(5)$.

\noindent\paragraph{More notation:}
In order to keep track of the various branches we will give the states and the moduli an index, e.g.\ $\psi_A$, $a_A, b_A, c_A$. Here $A=\alpha, \beta, \dots$ and for permutation points $A$ will denote the appropriate combination $A=\alpha\beta, \beta\gamma, \dots$. Later we will drop this index again as long as it is clear from the context which branch is being discussed.
\\

\subsection{Spectra}

Since we want to describe the moduli space it is enough to restrict to the marginal fermionic fields in the spectrum. The reason for this is that the moduli space is generated by deformations of the boundary BRST operator. Therefore we only need to be interested in fermions of charge one.

The spectrum at the permutation points is easily obtained from the cohomologies computed in section (\ref{MFintro}).
In (\ref{quintic_charges}) we give a list for the charges of each state in the cohomologies of the factors of $Q$ and indicate odd or even grading by the subscripts {\it f} or {\it b}.
\eqn{
\label{quintic_charges}
	(12): \quad& 0_b \quad {\tfrac{2}{5}}_b \quad {\tfrac{4}{5}}_b \quad {\tfrac{6}{5}}_b\\
	(34): \quad& 0_b \quad {\tfrac{2}{5}}_b \quad {\tfrac{4}{5}}_b \quad {\tfrac{6}{5}}_b\\
	(5): \quad	& 0_b \quad {\tfrac{3}{5}}_f \\
}
From this it is clear that one can find two fermions of charge 1. We write them symbolically as 
$\left[\frac{2}{5}\right]_b \odot [0]_b \odot \left[\frac{3}{5}\right]_f$ 
and $[0]_b \odot \left[\frac{2}{5}\right]_b \odot \left[\frac{3}{5}\right]_f$.
The numbers denote the R-charge and the subscript fermionic or bosonic grading.

In order to find the spectrum on the branch away from the permutation point it is necessary to explicitly compute the cohomology. The details of this calculation have been presented in \cite{Baumgartl:2007an}. It was shown that there are exactly two fermions present everywhere on the branch. One fermion is given by the derivative of the BRST charge with respect to a modulus, which we will denote by $\psi$. The second fermion we will denote by $\psi^\perp$. On the branch $(\alpha)$ we find the following explicit representation:
\eqn{
	\psi_\alpha &= \partial_{b_{\alpha}} Q^{\alpha} = -x_3 \pi_2 - \frac{b_\alpha^4}{c_\alpha^4}x_3 \pi_3 + \partial_b\left(E_2\bar\pi_2 + E_3\bar\pi_3\right)\\
	\psi_\alpha^\perp &=\frac{x_1}{x_3}\psi_\alpha\ .
}
There are two comments at order on these states: First, as mentioned above the matrix factorisation $Q$ from which the first fermion is derived, is not unique, but $Q$ is subject to a huge gauge symmetry, which is given by all transformations leaving $Q^2=W$ invariant. Therefore the explicit expressions given here are gauge dependent. 
The explicit form of $\psi_\alpha$ suggests that $x_3$ plays a distinguished role, but indeed it is possible to choose another equivalent $Q$ so that $\psi_\alpha$ is proportional to $x_4$ or $x_5$. This reflects democracy among the coordinates. The only restrictions come from the patch of $(a_\alpha:b_\alpha:c_\alpha)$ in which one is working; depending on that choice it is sometimes preferable to consider derivatives with respect to $a_\alpha$ or $c_\alpha$ rather than $b_\alpha$. This, of course, depends on to which permutation point one wants to connect the branch (\eg the permutation point $(\alpha\beta)$ cannot be described in the patch where $b_\alpha=1$).

Second, the same argument also applies to the coordinates $x_1$ and $x_2$. Hence it is feasible to switch to a matrix factorisation which comes with $x_1$ and $x_2$ exchanged. This also changes the relation between $\psi_\alpha$ and $\psi_\alpha^\perp $. Generally we can say: for a branch $(ij)(klm)$ it is possible to choose a factorisation so that its exactly marginal fermion is proportional either to $x_k$, $x_l$ or $x_m$. Furthermore it is possible to choose it in a way that the second fermion is proportional either to $x_i$ or $x_j$. This will be important when we join the branches together to form a web.

Before we look at the connections between the branches, we make a few comments on the nature of these fermions.

It is clear, by construction, that $\psi_\alpha$ is unobstructed on the branch since it creates the modulus. The second fermion $\psi_\alpha^\perp$ cannot be marginal, since we know that the moduli space is one-dimensional. This can be made explicit by computing the three-point function, which gives
\cite{Baumgartl:2007an, Kapustin:2003ga, Herbst:2004ax}
\eqn{
	\langle \psi_\alpha^\perp\psi_\alpha^\perp\psi_\alpha^\perp \rangle = -\frac{2}{5}\eta^4\frac{b_\alpha^3}{c_\alpha^9}\ .
}
Only at the point $b=0$ this fermion can become marginal, and this is consistent with the fact that there are two marginal fermions at a permutation point. When changing from one branch to the other, the two fermions exchange their roles, as has been shown in \cite{Baumgartl:2007an}.

\subsection{The moduli web}

We want to look a bit closer on what happens to the cohomology in the vicinity of a permutation point.
For example, at $(\alpha\beta)=(12)(35)(4)$ we can find two fermions
\eqn{
	f_{\alpha\beta}^1 = \one \odot x_3 \odot \begin{pmatrix} 0&1\\-x_4^3&0\end{pmatrix}
}
and 
\eqn{
	f_{\alpha\beta}^2 = x_1 \odot \one \odot \begin{pmatrix} 0&1\\-x_4^3&0\end{pmatrix}
}
On $(\alpha)$ we have (in the patch $a_\alpha=1$)
\eqn{
	\psi_\alpha &= \one \odot \partial_b Q^{\alpha} \\
	\psi_\alpha^\perp &= x_1 \odot \frac{1}{x_3}\partial_b Q^{\alpha}
}
so that it is obvious that $f_\alpha^1$ is simply the continuation of $\psi_\alpha$ at the permutation point, and likewise $\psi_\alpha^\perp$ becomes $f_\alpha^2$. 

On the branch $(\beta)=(35)(124)$ we find the two fermions
\eqn{
	\psi_\beta &= \one \odot \partial_b Q^{\beta} \\
	\psi_\beta^\perp &= x_3 \odot \frac{1}{x_1}\partial_b Q^{\beta}
}
By looking at their expressions at the permutation point $(\alpha\beta)$ one sees how the fermions on the branches can be identified:
\eqn{
	\psi_\alpha \sim &f_{\alpha\beta}^1 \sim \psi_\beta^\perp \\
	\psi_\alpha^\perp\sim &f_{\alpha\beta}^2 \sim \psi_\beta\ .
} 
The obstructed fermion in $(\alpha)$ becomes the unobstructed fermion on $(\beta)$ and vice versa. Only at the permutation point both fermions are marginal. So, locally in the vicinity of the permutation point, the moduli space is ${\mathbb C}\oplus{\mathbb C}$.

Let us now see how the branch $(\beta)$ connects $(\alpha\beta)$ with $(\beta\mu)=(35)(24)(1)$.
At $(\beta\mu)$ we find the marginal fermions
\eqn{
	f_\beta^1 = x_2 \odot \one \odot \begin{pmatrix} 0&1\\-x_1^3&0\end{pmatrix}
}
and 
\eqn{
	f_\beta^2 = \one \odot x_3 \odot \begin{pmatrix} 0&1\\-x_1^3&0\end{pmatrix}
}
Let us now consider $(\beta')=(35)(124)$, which differs from $(\beta)$ just by a gauge transformation. We find
\eqn{
	\psi_{\beta'} &= \one \odot \partial_b Q^{\beta'} \\
	\psi_{\beta'}^\perp &= x_3 \odot \frac{1}{x_2}\partial_b Q^{\beta}
}
On $(\mu)=(24)(135)$ we find the fermions
\eqn{
	\psi_{\mu} &= \one \odot \partial_b Q^{\mu} \\
	\psi_{\mu}^\perp &= x_2 \odot \frac{1}{x_3}\partial_b Q^{\beta}
}
So the fermions are connected in the following way:
\eqn{
	\psi_{\beta'} \sim &f_{\beta'\mu}^1 \sim \psi_\mu^\perp \\
	\psi_{\beta'}^\perp\sim &f_{\beta'\mu}^2 \sim \psi_\mu\ .
} 
Again we see the obstructed and the unobstructed fermion change their roles at the permutation point. Note that in order to see this it was important to correctly understand the appearance of the gauge transformation\footnote{The appearance of the gauge transformations as the price to pay that we work in coordinate patches where $a=1$ on each branch and consider $b$ as the free modulus.}.

With these preparations we can set up a chain of moduli branches: 
\eqn{
&\begin{matrix}(\alpha)\\(12)(435)\end{matrix}
\longrightarrow
	\begin{matrix}(\alpha\beta)\\(12)(35)(4)\end{matrix}
\longrightarrow
\begin{matrix}(\beta)\\(35)(412)\end{matrix}
\overset{\text{gt}}{\longrightarrow}
\begin{matrix}(\beta')\\(35)(124)\end{matrix}
\longrightarrow
\\
\longrightarrow
	&\begin{matrix}(\beta'\mu)\\(35)(24)(1)\end{matrix}
\longrightarrow
\begin{matrix}(\mu)\\(24)(135)\end{matrix}
\overset{\text{gt}}{\longrightarrow}
\begin{matrix}(\mu')\\(24)(315)\end{matrix}
\longrightarrow
	\begin{matrix}(\mu'\epsilon)\\(24)(15)(3)\end{matrix}
\longrightarrow
\\
\longrightarrow
&\begin{matrix}(\epsilon)\\(15)(324)\end{matrix}
\overset{\text{gt}}{\longrightarrow}
\begin{matrix}(\epsilon')\\(15)(234)\end{matrix}
\longrightarrow
	\begin{matrix}(\epsilon'\zeta)\\(24)(15)(3)\end{matrix}
\longrightarrow
\begin{matrix}(\zeta)\\(34)(215)\end{matrix}
\longrightarrow
\\
\overset{\text{gt}}{\longrightarrow}
&\begin{matrix}(\zeta')\\(34)(521)\end{matrix}
\longrightarrow
	\begin{matrix}(\zeta'\alpha)\\(34)(21)(5)\end{matrix}
\longrightarrow
\begin{matrix}(\alpha')\\(12)(534)\end{matrix}
\overset{\text{gt}}{\longrightarrow}
\begin{matrix}(\alpha)\\(12)(435)\end{matrix}
}

The arrows labelled by `gt' indicate a gauge transformation. 
This cycle $\alpha-\beta-\mu-\epsilon-\zeta-\alpha$ is not the only cycle we
can construct. The global structure of the moduli space is encoded in the
symmetries of the quintic and the intersections listed in table
\ref{branches}. In order to arrive at a convenient representation of the symmetries we
will map the data in table
1 to a graph by assigning a vertex to each branch and an edge to each
permutation point.
This graph is unoriented and contains self-intersecting faces.
We want to find an universal cover of this
graph which avoids such intersections and is oriented. The moduli space is then a quotient of it.

The smallest cycles one can find in this graph are cycles of length 5 and length 6, which define faces with 5 and 6 vertices. When orientation is taken into account we find 12 such cycles of length 5 and 20 cycles of length 6. The maps from $S^2$ to graphs which consist only out of pentagons and hexagons have been classified in \cite{Braungart:2006}. The minimal standard realisation is the uniform polyhedron $U_{25}$, the truncated icosahedron also known as soccer ball.

\begin{figure}
\centering
\includegraphics[width=140mm, viewport=100 260 550 650]{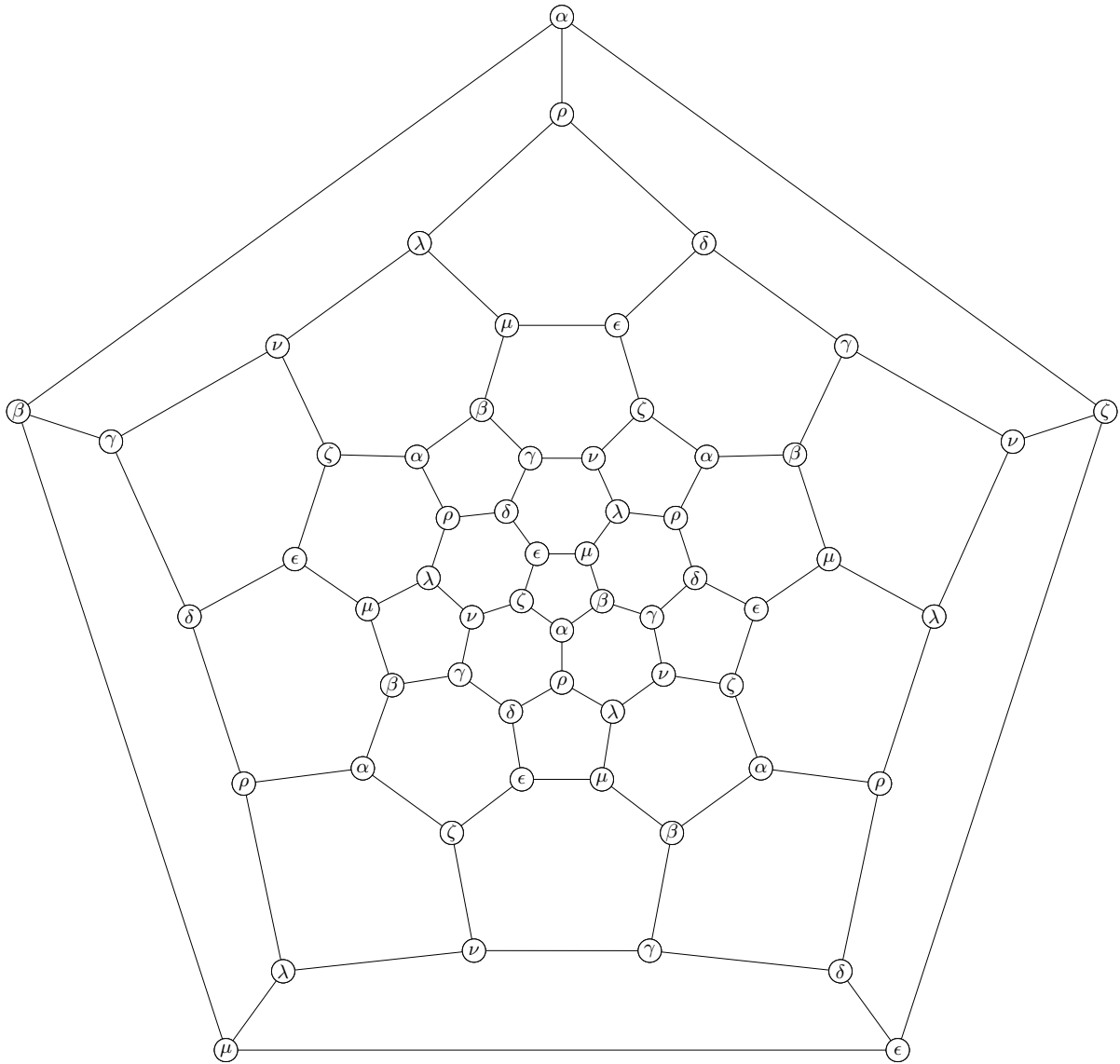}
\caption{The Schlegel tree diagram for the truncated icosahedron $U_{25}$. Vertices represent moduli branches and are labelled according to table \ref{branches}. Vertices with same label are identified. Edges correspond to intersections of branches and are therefore identified with permutation points.}
\label{schlegel1}
\end{figure}

Figure \ref{schlegel1} shows the Schlegel tree diagram associated to $U_{25}$. The vertices have been decorated with the names of the moduli branches which they represent. Each vertex appears 6 times, so the topology of the moduli space must be a quotient of $U_{25}$. The automorphism group of the `soccer tree' has been determined in \cite{Braungart:2006} as $\text{Aut}(U_{25})={\mathbb Z}_2\times {\mathbb Z}_3 \times {\mathbb Z}_5$. However, this is not the automorphism group which we encounter for the soccer ball with labelled vertices. Rather, we are missing the isotropy group of order 5 in \cite{Braungart:2006} for the pentagons, which leads to an automorphism group of ${\mathbb Z}_2\times {\mathbb Z}_3$ in our case.
Therefore we find that the moduli space of D5-branes on the undeformed quintic has the symmetry\footnote{If we consider the ${\mathbb Z}_5$-orbifold of the Landau-Ginzburg theory, then indeed the ${\cal M}_{D5} \simeq \frac{U_{25}}{ {\mathbb Z}_2\times {\mathbb Z}_3 \times {\mathbb Z}_5}$.}
\eqn{
	{\cal M}_{D5} \simeq \frac{U_{25}}{ {\mathbb Z}_2\times {\mathbb Z}_3}\ .
}

In the following sections we will discuss examples of other threefolds. It will become obvious that their moduli space is given by the same web as the Fermat quintic. The construction applies to all threefolds constructed out of tensor products of minimal models.

Before we do this we want to discuss superpotentials on this web.

\subsection{Bulk deformations and superpotential}

Non-vanishing bulk-boundary correlators contain information on the effective superpotential. This tells us which directions are flat and which are possibly lifted under bulk deformations. We have seen above that in the presence of bulk operators the boundary moduli space collapses into a set of discrete points. Only these points preserve supersymmetry and are obtained as extrema of a bulk induced potential. This connection has been investigated in \cite{Baumgartl:2007an}.

With our approach we are in the convenient situation that we know the boundary moduli space {\em exactly}, therefore we can study effects of bulk perturbations globally. By integrating three-point functions we are able to determine the effective superpotential ${\cal W}$ explicitely. On each of the branches the bulk-boundary couplings satisfy
\eqn{
\label{eqRG}
	\partial_b{\cal W} =\frac{\lambda}{2} B_{G\psi}\ ,
}
where $b$ is the coupling associated to the boundary fermion $\psi$ and $G$ is a bulk operator.
A closed expression for the superpotential can only be obtained because the bulk-boundary correlators are known on every point of the moduli space and can therefore be integrated up.

We will deform the superpotential by $\Delta W=G$, an element of the bulk chiral ring:
\eqn{
	W'=W+G .
}
The deformation we are interested in is of the form
\eqn{
\label{bulkdef}
	G &= \lambda s^{(3)}(x_1, x_2) s^{(2)}(x_3, x_4, x_5)\ ,
}
where $s^{(n)}(x_1, x_2, \dots, x_k) = \sum_{n=\sum r_i} s^{(n)}_{r_1r_2\dots r_k} x_1^{r_1}x_2^{r_2}\cdots x_k^{r_k}$ denotes homogenous polynomials of degree $n$.

These monomials have a non-vanishing bulk-boundary correlator with the fermion $\psi_1$, which generates the branch $(\alpha)$. There is a second disjoint class of bulk deformations,
\eqn{
	H= s^{(2)}(x_1, x_2) s^{(3)}(x_3, x_4, x_5)\ ,
}
which excite the completely obstructed fermion $\psi_\alpha^\perp=\frac{x_1}{x_3}\psi_\alpha$ on $(\alpha)$.

On the quintic perturbed by $G$ the matrix factorisation condition can only be satisfied for a set of discrete points on the branch. This set has been shown to be determined by the intersection of the curves $a^5+b^5+c^5=0$ and $s^{(2)}(a,b,c)=0$ at exactly 10 points\footnote{For the sake of readability we will not indicate the branch which some moduli are associated to as long as this is clear from the context. Otherwise, for instance $a^{(\alpha)}$ will indicate a modulus on $(\alpha)$.}. At these points, the bulk deformation need not be infinitesimal, but matrix factorisations can be constructed also for finite $\lambda$.

From symmetry considerations it is easy to determine, which fermions on other branches are excited. For example, $G=x_1^2x_2x_3x_4$ will give a potential to the branch generating fermions on $(\alpha)$, $(\gamma)$ and $(\lambda)$.

The effective superpotential is obtained by integrating the bulk-boun\-dary correlators with the fermion which generates the branch. This is possible because the correlators are holomorphic functions on the moduli space \cite{Baumgartl:2007an}.
The correlation functions obtained are
\eqn{
	\vev{s^{(3)}(x_1,x_2)\cdot s^{(2)}(x_3, x_4, x_5) \cdot \psi} 
		&= -\frac{\eta}{25}s^{(3)}(\eta,1) \frac{s^{(2)}(1,b,c)}{c^4} \\
	\vev{s^{(2)}(x_1,x_2)\cdot s^{(3)}(x_3, x_4, x_5) \cdot \psi^\perp} 
		&= -\frac{\eta^2}{25}s^{(2)}(\eta,1) \frac{s^{(3)}(1,b,c)}{c^4}\ .
}
In particular there is a one-to-one map between the coefficients of monomials in $s^{(2)}$ and globally holomorphic forms on the Riemann curve $a^5+b^5+c^5=0$ \cite{Baumgartl:2007an}. Thus equation (\ref{eqRG}) can be integrated. The superpotentials we obtain this way are given in terms of hypergeometric functions by (we have skipped some unimportant global prefactors)
\eqn{
	{\cal W}(1,b,c)=\lambda  \sum_{i+j+k=2} s^{(2)}_{ijk}{\cal W}_{j+1,k+1}
}
on the branch $(\alpha)$ in the patch where $a=1$. 
Here (and in the following)
\eqn{
	{\cal W}_{rs}&=\frac{b^r}{r}{}_2{\sf{F}}_1\left(\frac{r}{N},1-\frac{s}{N};1+\frac{r}{N};-{b}^{N}\right),\qquad
	N=5\ .
}

\section{The Calabi-Yau {$\mathbb P_{(1,1,1,1,2)}[6]$}}
\label{sec:P6}

In this section we investigate the moduli space of D5-branes in $\mathbb P_{(1,1,1,1,2)}[6]$ with the methods developed for the Fermat quintic in the previous section.
This Calabi-Yau manifold is defined by the superpotential
\eqn{
\label{eq:W1}
  W=x_1^6+x_2^6+x_3^6+x_4^6+x_5^3
}
of weighted projective degree 6 \cite{Candelas:1990rm, Klemm:1992tx}. In particular the field $x_5$ has charge $\frac{2}{3}$ while all other fields $x_i$ carry charge $\frac{1}{3}$. 

Anticipating our results, we will find the same structure for the moduli web, but we will see that there are now additional obstructed fermions at some of the permutation points.

\subsection{Embedded lines}

For the
superpotential $W$ a parametric equation for (families of) lines is
\eqn{
\label{CY2lineA}
  \ell_1=(u:\eta u:av:bv:cv^2)
}
where $(u:v)\in \mathbb{P}^1$ and $\eta$, $a$, $b$, $c$ have to be chosen such that
$(\alpha)$ lies in the Calabi-Yau. In other words they have to satisfy the equation
\eqn{
  u^6+\eta^6 u^6+a^6v^6+b^6v^6+c^3v^6=u^6(1+\eta^6)+v^6(a^6+b^6+c^3)=0.
}
Therefore $\eta$ is a sixth root of $-1$ and $a,b,c$ satisfy
\eqn{
  a^6+b^6+c^3=0 \subset  \mathbb{P}_{[1,1,2]}\ .
}
Thus $(\alpha)$ is parametrised by a Fermat curve.

There is a second inequivalent type of lines in $W$, namely those where $x_5$ is parametrised by the coordinate $u$:
\eqn{
\label{CY2lineB}
  \ell_2=(u:av:bv:cv:\eta^2u^2)\ .
}
The condition on the moduli is now
\eqn{
	a^6+b^6+c^6=0 \subset  \mathbb{P}^2
}for these lines.

\subsection{Permutation points}

The intersection pattern of lines does not differ from the Fermat quintic case, since we are still considering a situation where five minimal models are tensored. Therefore the soccer ball diagram is valid here, too, and can be used to keep track of all the permutation points and moduli branches. 

The matrix factorisations associated to line of type (\ref{CY2lineA}) are given by
the branches which we call of `type 1' $(\alpha), (\gamma), (\delta), (\zeta), (\lambda)$ and $(\mu)$. The other branches  which we call of `type 2'
$(\beta), (\epsilon), (\nu)$ and $(\rho)$ correspond to lines of the type (\ref{CY2lineB}).
We find permutation points that correspond to intersections of lines of the first type and those that are intersection between first and second type, but no intersections between second type lines only. 

In this setting we find two types of permutation points. First, there is
\eqn{
	(i j)(k l)(5)\ ,
}
which is an intersection between two lines of first type. The matrix factorisation is determined by
\eqn{
	J_1 &= x_i-\eta x_j \\ 
	J_2 &= x_k-\eta' x_l\\
	J_3 &= x_5\ .
}
The states in the cohomologies of the factors $Q_i$ are listed by charge:
\eqn{
\label{CY2chargesAA}
(ij):\quad & 0_b \quad {\tfrac{1}{3}}_b \quad {\tfrac{2}{3}}_b \quad {1}_b \quad {\tfrac{4}{3}}_b\\
(kl):\quad & 0_b \quad {\tfrac{1}{3}}_b \quad {\tfrac{2}{3}}_b \quad {1}_b \quad {\tfrac{4}{3}}_b\\
 (5):\quad & 0_b \quad {\tfrac{1}{3}}_f 	
}
From this we get three marginal fermions:
\eqn{
\label{CY2threeF}
	f_1 &= [0_b]\odot \left[{\tfrac{2}{3}}_b\right] \odot \left[{\tfrac{1}{3}}_f\right] \\
	f_2 &= \left[{\tfrac{2}{3}}_b\right] \odot [0_b] \odot \left[{\tfrac{1}{3}}_f\right] \\
	f_3 &= \left[{\tfrac{1}{3}}_b\right] \odot \left[{\tfrac{1}{3}}_b\right] \odot \left[{\tfrac{1}{3}}_f\right] \\
}

An example for the other class of permutation points, which is an intersection of a type 1 and a type 2 branch, is given by the matrix factorisation
$(ij)(l5)(k)$
with the polynomials
\eqn{
	J_1 &= x_i-\eta x_j \\ 
	J_2 &= x_l^2-\eta'^2 x_5\\
	J_3 &= x_k\ .
}
Note that this is a product of a permutation brane $(ij)$, a generalised permutation brane $(l5)$ and a minimal model $(k)$.
The states are:
\eqn{
\label{CY2chargesAB}
(ij):\quad& 0_b \quad {\tfrac{1}{3}}_b\quad {\tfrac{2}{3}}_b \quad {1}_b \quad {\tfrac{4}{3}}_b\\
(l5):\quad& 0_b \quad {\tfrac{1}{3}}_b \quad {\tfrac{2}{3}}_b\quad {1}_b \\
(k):\quad& 0_b \quad{\tfrac{2}{3}}_f 
}
The marginal fermions are:
\eqn{
	f_1 &= [0_b]\odot \left[{\tfrac{1}{3}}_b\right] \odot \left[{\tfrac{2}{3}}_f\right] \\
	f_2 &= \left[{\tfrac{1}{3}}_b\right] \odot [0_b] \odot \left[{\tfrac{2}{3}}_f\right] \\
}
Note that there are no intersections between two type 2 branches.

\subsection{Marginal cohomology on the branches}
\label{CY1margcoh}

As next set we want to determine the marginal fermions away from the permutation points. 
It is clear immediately that the single fermion which generates a branch is given by a derivative of $Q$ with respect to a modulus, since $\{ Q, \partial_b Q\} = \frac{1}{2}\partial_b\{ Q, Q\} = \frac{1}{2}\partial b W = 0$.

Let us consider the branch $(ij)(kl5)$. The corresponding matrix factorisation is given by
\eqn{
	J_1&=x_i-\eta x_j\\
	J_2&=ax_k-bx_l\\
	J_3&=cx_l^2-a^2x_5\ .
}
The exactly marginal fermion $\psi=\partial_b Q$ is obviously proportional to $x_l$. 
Therefore we can construct a second state of weight 1 by dividing out $x_l$ and replacing it with $x_i$. This state must be $Q$-closed since $\{ Q, \frac{x_i}{x_l}\partial_b Q\} = \frac{1}{2}\frac{x_i}{x_l}\partial_b\{ Q, Q\} = \frac{1}{2}\frac{x_i}{x_l}\partial b W = 0$\footnote{Instead of $\frac{x_i}{x_l}$ we could have considered polynomials $f(x_1,x_2,x_3,x_4,x_5)$ of weight 1 (this corresponds to degree 1 only for the first four coordinates). A choice of $x_3, x_4$ or $x_5$ gives a state in the same equivalence class as $\psi$. A choice of $x_1$ or $x_2$ gives the different state $\psi^\perp$.}.
We thus get the states
\eqn{
\label{CY2fermionsA}
	\psi &= \partial_b Q \\
	\psi^\perp &= \frac{x_i}{x_l}\partial_b Q\ .
}
This is in fact only true when we are away from certain permutation points. For example, when we approach the point given by $b=0$ we see that $\partial_b Q$ becomes now proportional to $x_l^2$. This shows that at such a point we can construct the states $\frac{x_i}{x_l}\partial_bQ(b=0)$ and $\frac{x_i^2}{x_l^2}\partial_bQ(b=0)$. Of course this is what we find from the examination of the brane configuration at the permutation point. Away from the permutation point we can only find two fermions with this method, and indeed an explicit calculation presented in appendix \ref{CY1branches} proves that this is the full cohomology.

The dimension of the marginal cohomology as computed in appendix \ref{CY1branches} is the generic dimension on the Riemann curve which forms the moduli space. The linear system of equations from which the cohomology is calculated parametrically depends on the position in the curve. As is explained in the appendix, there might be special points at which the cohomology jumps, but if so the dimension will be larger than on generic points. We can utilise this fact and check if we have found the full cohomology on a branch from computing the cohomologies at the permutation points. For example, we know that the branches of type 1 and 2 intersect in a permutation point whose cohomology is two-dimensional. Since we have found already two fermions on branch 1 we can be sure that we have found all marginal fermions.

The same argument also applies to the second branch, where we also expect 2 or less fermions. On branches $(5j)(klm)$ we work with the matrix factorisation
\eqn{
\label{CY2fermionsMF}
	J_1 &= x_5 - \eta^2 x_j \\
	J_2 &= ax_k - bx_l \\
	J_3 &= cx_l - a x_m\ .
}
Just as before we construct the two fermions
\eqn{
\label{CY2fermionsB}
	\psi &= \partial_b Q \\
	\psi^\perp &= \frac{x_5}{x_l}\partial_b Q\ .
}
With the arguments presented above we have constructed the full cohomology.

\subsection{The moduli web}

As we have pointed out above the tensor structure of the model consisting of five minimal models makes it obvious that the various branches of moduli form the soccer ball diagram as in the case of the Fermat quintic. We must check though if the cohomologies can really be joined together at the permutation points.

Let us first consider the permutation point between branches of first and second type, 
$Q=(ij)(kl5)$ and $Q'=(5l)(ijk)$ with intersection $(ij)(5l)(k)$.
On $Q$ we find the fermions $\psi=\partial_b Q$ and $\psi^\perp=\frac{x_j}{x_l}\partial_b Q$. 
On $Q'$ we have $\psi'=\partial_b Q'$ and ${\psi'}^\perp=\frac{x_l}{x_j}\partial_b Q'$. Here we have already chosen a gauge in which it is obvious that at the permutation point the two fermions are exchanged
\eqn{
	\psi^{\phantom{\perp}}&\longleftrightarrow {\psi'}^\perp\\ 
	\psi^\perp &\longleftrightarrow \psi'\ .
}

Two branches of first type 
$Q=(ij)(5lm)$ and $Q'=(lm)(5ji)$ intersect at $(ij)(lm)(5)$.
On $Q$ live the fermions $\psi=\partial_b Q$ and $\psi^\perp=\frac{x_j}{x_l}\partial_b Q$. 
On $Q'$ there are $\psi'=\partial_b Q'$ and ${\psi'}^\perp=\frac{x_l}{x_j}\partial_b Q'$. 
Thus there are two fermions present on each branch, but at the permutation point itself the marginal fermionic cohomology is enhanced and consists of three states (\ref{CY2threeF})
\eqn{
	f_1 &= x_l^2\left(\pi^3 - x_5\bar\pi^3\right) \\
	f_2 &= x_j^2\left(\pi^3 - x_5\bar\pi^3\right) \\
	f_3 &= x_jx_l\left(\pi^3 - x_5\bar\pi^3\right)\ .
}
Transporting the fermions from the branches to the permutation point yields the connections
\eqna{
	&\psi^{\phantom{\perp}} \longrightarrow &f_1  \nonumber\\
	&\psi^\perp  \longrightarrow &f_3  \longleftarrow {\psi'}^\perp \\
	&&f_2\longleftarrow \psi' \ .\nonumber
}
In particular we observe here that the obstructed fermion on both branches can be identified. At the permutation point the branch generating fermions appear in or disappear from, respectively, the marginal cohomology.

\subsection{Obstructions}

In this section we compute three-point functions of the fermions. Let us first focus on the branches of second type, $(i5)(klm)$.
We find the following correlators:
\eqn{
	\langle (\psi)^3\rangle &= 0\\
	\langle ({\psi^\perp})^3\rangle &\propto \frac{b^4}{c^{11}}\\
}
This shows that $\psi^\perp$ is obstructed to lowest order everywhere except at the permutation point $b=0$, where the transition to the next branch occurs.

On the branches of first type, $(ij)(kl5)$ the correlation functions all vanish
\eqn{
	\langle (\psi)^3\rangle &= 0\\
	\langle ({\psi^\perp})^3\rangle &=0\\
}
(and this implies the vanishing of the three-point functions of all $f_i$ at the permutation point, too).
From our above the field $\psi^\perp$ is supposed to be obstructed, since our geometric picture tells us that it does not generate a moduli branch. As its three-point function vanishes we expect the obstructions to occur at higher order.

In order to see this we perturb the BRST operator by $\psi^\perp$ and apply the methods developed in \cite{Hori:2004ja}.
For convenience we present here the argument only at the permutation point and refer to appendix \ref{CY1obstr} for the general calculation.
Our ansatz is
\eqn{
	Q(\lambda) = \sum_n \lambda^nQ_n\ ,
}
where $Q_0$ is the original $Q$ and $Q_1=\psi^\perp$. The first order equation $\{Q_0,Q_1\}=0$ is satisfied because $\psi^\perp$ is in the cohomology. The second order equation, which determines $Q_2$, is
\eqn{
	\{ Q_0, Q_2\} = -\frac{1}{2} \{ Q_1, Q_1\} \propto x_j^2x_l^2x_5\ .
}
Indeed we can find an appropriate $Q_2$, e.g.\
\eqn{
	Q_2 &= x_j^2x_l^2\bar\pi^3\ .
}
The third order equation is
\eqn{
	\{ Q_0, Q_3\} = -\{Q_1, Q_2\} \propto x_j^3 x_l^3\ .
} 
At this order the perturbation series breaks down because the r.h.s. is not $Q$-exact, and therefore no $Q_3$ can be found. We encounter an obstruction at third order in $\lambda$.

\subsection{Bulk deformations}

Under general bulk deformations not all branes on the moduli space will stay supersymmetric. We want to identify those bulk deformations for which we can find branes whose moduli space extends into a bulk direction. Since we can probe only the on-shell properties of these branes, the boundary modulus will be fixed. 

For the branches of first type $(ij)(kl5)$ we find the following bulk boundary correlators:
\eqn{
	\left\langle s^{(4)}(x_i, x_j) s^{(2)}(x_l, x_k, x_5) \cdot \psi\right\rangle 
		&= -\frac{\eta s^{(4)}(\eta,1) }{18}\frac{s^{(2)}(1, b, c)}{c^2} \\
	\left\langle s^{(3)}(x_i, x_j) s^{(3)}(x_l, x_k, x_5) \cdot \psi^\perp\right\rangle 
		&= -\frac{\eta^2 s^{(3)}(\eta, 1)}{18} \frac{s^{(3)}(1, b, c)}{c^2} \\
}
while for the second type branches $(i5)(klm)$ we get
\eqn{
	\left\langle s^{(3)}(x_i, x_5) s^{(3)}(x_l, x_k, x_m) \cdot \psi\right\rangle 
		&= \frac{\eta s^{(3)}(\eta,1)}{18} \frac{s^{(3)}(1, b, c)}{c^5} \\
	\left\langle s^{(2)}(x_i, x_5) s^{(4)}(x_l, x_k, x_m) \cdot \psi^\perp\right\rangle 
		&= \frac{\eta s^{(2)}(\eta,1)}{18} \frac{s^{(4)}(1, b, c)}{c^5} \\
}
Here, $s^{(n)}(x_{j_1},x_{j_2}, \dots, x_{j_m})=\sum_{i_1+\dots+i_m=n} s^{(n)}_{i_1\dots i_m} x_{j_1}^{i_1}\cdots x_{j_m}^{i_m}$ are homogenous polynomials of weighted degree $n$.

In order to find branes which deform with a bulk deformation they must satisfy
\eqna{
	&a^6+b^6+c^3 = 0 = s^{(2)}(a,b,c) \qquad &\text{for type 1} \\
	&a^6+b^6+c^6 = 0 = s^{(3)}(a,b,c) \qquad &\text{for type 2} \ .
}
There are 12 and 18 such points on a branch. These points are determined by the chosen bulk deformation. 
The associated matrix factorisations are given as follows:

The bulk deformation is
\eqn{
	W \to W' = W + G
}
where
\eqna{
	G =& \lambda s^{(4)}(x_i, x_j) s^{(2)}(x_k, x_l, x_5) \qquad &\text{ for type 1} \\
	G =& \lambda s^{(3)}(x_i, x_5) s^{(3)}(x_k, x_l, x_m) \qquad &\text{ for type 2} \ .
}
The deformations of the matrix factorisations are given by
\eqn{
	\text{type 1:} &\\
	E_2 &\to E_2 + \lambda s^{(4)}(1,\eta)\left(-\frac{a^2 s^{(2)}_{200} + cs^{(2)}_{001}}{a^2b}x_l + \frac{s^{(2)}_{020}}{a}x_k\right) \\
	E_3 &\to E_3 - \lambda s^{(4)}(1,\eta)\frac{s^{(2)}_{001}}{a^2}
}

\eqn{
	\text{type 2:}& \\
	E_2 &\to E_2 + \lambda s^{(3)}(\eta^2,1)\Bigl(
		x_k^2 \frac{s_{030}}{a} +x_m^2 \frac{s_{012}-s_{021}}{a} -x_k x_m \frac{a s_{111} +c s_{021}}{b a}\\
&\quad-x_l^2 \frac{a^3 s_{300}+a^2c s_{201} +ac^2 s_{102}+bc^2 s_{012}-bc^2 s_{021}+c^3 s_{003}}{a^3 b}\\
&\quad-x_k x_l \frac{a^2c s_{201} +ac^2 s_{102} +bc^2  s_{012}-bc^2  s_{021}+c^3 s_{003}+ a^2bs_{210}+a^3 s_{300}}{b^2 a^2}\Bigr)\\
	E_3 &\to E_3 + \lambda s^{(4)}(\eta^2,1)\Bigl(
	-x_m^2 \frac{s_{003}}{a}-x_k x_m \frac{s_{021}}{a}\\
&\quad     -x_l x_m \frac{as_{102} +b s_{012}-b s_{021}+c s_{003}}{a^2}\\
&\quad		x_k^2 \frac{a b^2s_{120} +a^2c s_{201}+ac^2 s_{102} +bc^2  s_{012}-bc^2  s_{021}}{ab^2 c } \\
&\quad		x_k^2 \frac{c^3 s_{003}+a^2bs_{210}+a^3 s_{300}+b^3 s_{030}}{ab^2 c } \\
&\quad-x_l^2 \frac{ a^2s_{201}+ac s_{102} +bc  s_{012}-bc  s_{021}+c^2 s_{003}}{a^3}
	\Bigr)
}
Those branes which do not lie at the points $s^{(2)}(a,b,c) = 0$ or $s^{(3)}(a,b,c)=0$, respectively, cannot be deformed by bulk fields. For them, supersymmetry is broken, which will render them instable. For the quintic it was possible to derive an effective superpotential for the brane moduli.

\subsection{Effective superpotentials}

The effective superpotential for bulk and boundary moduli obeys
\eqn{
	\partial_b {\cal W}_{\text{eff}} =\frac{\lambda}{2} B_{G\psi}\ ,
}
where the right hand side is given by a bulk-boundary correlator.
A closed expression for the superpotential can only be obtained when $B_{G\psi}$ is known on every point of the moduli space and can thus be integrated up.

In \cite{Baumgartl:2007an} it was a crucial observation that the bulk-boundary correlators form a set of holomorphic functions on the complete moduli space. This assigns a very concrete geometrical meaning to $B_{G\psi}$ and  is in fact the decisive criterion which tells us that the bulk-boundary couplings are derivatives of an effective potential. 

For the moduli branches of type 2 a basis for the bulk-boundary correlators $c^{-5}s^{(3)}(1, b, c)$ are the functions
\eqn{
	\left\{ \frac{b^rc^{s}}{c^5}, 0\le r+s \le 3\right\}\ .
}
This is in fact the complete set of holomorphic functions on the Riemann curve $1+b^6+c^6=0$ \cite{Griffiths}. 
The basis is 10-dimensional and this is also the genus $g=10$ of the curve. 
Therefore the associated holomorphic 1-forms are integrable. Their integrals are
\eqn{
	{\cal W}_{rs}=\frac{b^r}{r}{}_2{\sf{F}}_1\left(\frac{r}{6},1-\frac{s}{6};1+\frac{r}{6};-{b}^{6}\right)\ .
}
As result we get an expansion of the effective potential in terms of hypergeometric functions (we have ignored global factors which do not depend on the moduli)
\eqn{
  \mathcal{W}_{\text{eff}}^{\text{type 2}}
  	=& \sum_{i+j+k=3}s^{(3)}_{ijk} 
			{\cal W}_{j+1, k+1}
}

For the type 1 lines only a subset of the holomorphic functions appears as basis for the bulk-boundary correlators. This is due to the fact that fields of different weights appear in the perturbing polynomial $s^{(2)}(x_l, x_k, x_5)$. The basis is explicitly
\eqn{
	\left\{ \frac{b^ic^{j}}{c^2}, 0\le i+2j \le 2\right\}\ .
}
From this we see that the genus of the curve is $g=4$.
For the superpotential we find
\eqn{
  \mathcal{W}_{\text{eff}}^{\text{type 1}} = \sum_{i+j+2k=2}
  		s^{(2)}_{ijk}
		{\cal W}_{j+1, k+1}
}

\section{The Calabi-Yau {$\mathbb P_{(1,1,1,1,4)}[8]$}}
\label{CY2}

D5-branes on the 3-fold defined by
\begin{align}
  W=x_1^8+x_2^8+x_3^8+x_4^8+x_5^2=0
\end{align}
is technically very similar to the case discussed in the previous section. The main difference is that the weight of the coordinate $x_5$ is much larger in this example.

\subsection{Lines and the moduli web}

We find again two types of lines with the parametric equations
\eqn{
\label{CY3lineA}
  \ell_1=(u:\eta u:av:bv:cv^4)
}
and
\eqn{
\label{CY3lineB}
  \ell_2=(u:av:bv:cv:\eta^4u^4)
}
with the moduli spaces
\eqn{
	\eta^8&=-1\qquad a^8+b^8+c^2=0\qquad\subset{\mathbb P_{(1,1,4)}}\qquad \text{(type 1)} \\
	\eta^8&=-1\qquad a^8+b^8+c^8=0\qquad\subset{\mathbb P^2}\qquad\qquad\!\text{(type 2)} \\
}

At those permutation points $(ij)(kl)(5)$ which join two type 1 branches, we find the marginal fermions
\eqn{
	f_1 &= x_j^4(\pi^3-\bar\pi^3) \\
	f_2 &= x_j^3x_l(\pi^3-\bar\pi^3) \\
	f_3 &= x_j^2x_l^2(\pi^3-\bar\pi^3) \\
	f_4 &= x_jx_l^3(\pi^3-\bar\pi^3) \\
	f_5 &= x_l^4(\pi^3-\bar\pi^3)\ .
}
The marginal fermions at the other permutation points $(ij)(l5)(k)$ are
\eqn{
	f_1 &= x_l(\pi^3-x_k^6\bar\pi^3) \\
	f_2 &= x_j(\pi^3-x_k^6\bar\pi^3) \\
}

We now describe the fermions on the branches. Both on type 1 branches $(ij)(kl5)$ 
and on type 2 branches $(5j)(klm)$ we find
\eqn{
	\psi &= \partial_b Q \\
	\psi^\perp &= \frac{x_j}{x_l}\partial_b Q\ .
}
Both branches intersect in a permutation point whose marginal cohomo\-logy is two-dimensional. From the arguments presented in section \ref{CY1margcoh} it is clear that this is the full cohomology.

The joining relations at the permutation points are given by
\eqna{
	&\psi^{\phantom{\perp}} \longrightarrow &f_5  \nonumber\\
	&\psi^\perp  \longrightarrow &f_4  \nonumber\\
	&&f_3\\
	&& f_2 \longleftarrow {\psi'}^\perp \nonumber\\
	&& f_1\longleftarrow \psi' \ .\nonumber
}
for the intersections $(ij)(kl)(5)$, and
\eqna{
	&\psi^{\phantom{\perp}} \longrightarrow &f_1  \longleftarrow {\psi'}^\perp \nonumber\\
	&\psi^\perp  \longrightarrow &f_2  \longleftarrow \psi' 
}
for the intersections $(ij)(l5)(k)$.
The fermion $\psi^\perp$ on branch 2 has a three-point function
\eqn{
	\<(\psi^\perp)^3\> = \frac{7}{4}\eta^4\frac{b^6}{c^{15}}\ .
}
As expected, it is obstructed everywhere except at the permutation points.

On the first branch, $\psi^\perp$ has a vanishing three-point function. Also, the three-point function for the fermion $f_3$ at the permutation point vanishes. Again, we expect the obstructions to appear at higher order. It will be enough to check this at the permutation point, \ie for the fermions $f_2$, $f_3$ and $f_4$. For the first order perturbation we find
\eqn{
	-\frac{1}{2}\{f_2,f_2\} = x_j^6x_l^2\qquad
	-\frac{1}{2}\{f_3,f_3\} = x_j^4x_l^4\qquad
	-\frac{1}{2}\{f_4,f_4\} = x_j^2x_l^6\ .
}
All these expressions are non-trivial in cohomology, so a solution to the first order equations cannot be found. Thus these fermions are obstructed at first order.

\subsection{Bulk perturbations and effective superpotentials}

When switching on bulk moduli we find the following bulk-boundary correlators on branches of the first type:
\eqn{
\label{P8bb1}
	\< s^{(6)}(x_i, x_j)\cdot s^{(2)}(x_k, x_l, x_5) \cdot \psi\> &=  -\frac{\eta s^{(6)}(\eta,1)}{16} \frac{s^{(2)}(1, b, c)}{c} \\
	\< s^{(5)}(x_i, x_j)\cdot s^{(3)}(x_k, x_l, x_5) \cdot \psi^\perp\> &=  -\frac{\eta^2 s^{(5)}(\eta, 1)}{16} \frac{s^{(3)}(1, b, c)}{c}\ .
}
On type 2 branches we find
\eqn{
\label{P8bb2}
	\< s^{(3)}(x_i, x_5)\cdot s^{(5)}(x_k, x_l, x_m) \cdot \psi\> 
		&=  \frac{\eta s^{(3)}(\eta,1)}{16} \frac{s^{(5)}(1, b, c)}{c^7} \\
	\< s^{(2)}(x_i, x_5)\cdot s^{(6)}(x_k, x_l, x_m) \cdot \psi^\perp\> 
		&=  \frac{\eta^2 s^{(2)}(\eta, 1)}{16} \frac{s^{(6)}(1, b, c)}{c^7}\ .
}
(Note that in (\ref{P8bb1}) and (\ref{P8bb2}) the polynomials $s^{(2)}$ and $s^{(3)}$ are independent of $x_5$ due to its high charge. Hence also on the r.h.s.\ there is no dependence on $c$ or $\eta$.)

From this we can immediately derive the intersection points for those branes which deform under finite bulk deformations. They are given by 16 points $a^8+b^8+c^2=0=s^{(2)}(a,b,c)$ for the type 1 branches, and by 40 points $a^8+b^8+c^8=0=s^{(5)}(a,b,c)$ for the type 2 branches.

The basis of functions on the moduli space which is spanned by $\frac{s^{(5)}(1, b, c)}{c^7}$ in (\ref{P8bb2}) is, as in the examples above, in one-to-one correspondence to holomorphic differentials  \cite{Griffiths}
\eqn{
	\left\{\frac{b^rc^s}{c^7}, 0\le r+s\le 5\right\}
}
on the Riemann curve $a^8+b^8+c^8=0$. Its genus is $g=21$. The moduli space $a^8+b^8+c^2=0$ for the type 1 lines has genus $g=3$. The differentials are 
\eqn{
	\left\{\frac{b^rc^s}{c}, 0\le r+4s\le 2\right\}
}
and this is in clear correspondence to the r.h.s.\ of (\ref{P8bb1}). The integrated effective potential is given by
\eqn{
  \mathcal{W}_{\text{eff}}^{\text{type 1}}
  	=& \lambda s^{(6)}(1,\eta)\sum_{i+j+4k=2}s^{(2)}_{ijk} {\cal W}_{j+1, 4(k+1)}
}
and
\eqn{
  \mathcal{W}_{\text{eff}}^{\text{type 2}}
  	=& \lambda s^{(3)}(1,\eta^4)\sum_{i+j+k=6}s^{(5)}_{ijk} {\cal W}_{j+1, k+1}
}
where
\eqn{
	{\cal W}_{rs}=\frac{b^r}{r}{}_2{\sf{F}}_1\left(\frac{r}{8},1-\frac{s}{8};1+\frac{r}{8};-{b}^{8}\right)
}

\section{The Calabi-Yau {$\mathbb P_{(1,1,1,2,5)}[10]$}}
\label{CY3}

The last Calabi-Yau manifold we want to investigate is given by the defining equation
\eqn{
	W=x_1^{10}+x_2^{10}+x_3^{10}+x_4^5+x_5^2=0
}
in ${\mathbb P}_{(1,1,1,2,5)}[10]$.
We can geometrically embed the following types of lines into $W=0$:\\[8pt]
\begin{tabular}{|c|c|c|c|}
\hline
	&line parametrisation & factorisation& branches \\
\hline
  type 1&$\ell_1=( u:  \eta u: av: b v^2: c v^5)$ &(ij)(4l5) & $(\alpha), (\delta), (\lambda)$\\
  type 2&$\ell_2=(u: av: bv: \eta^2 u^2: c v^5)$ &(i4)(kl5) & $(\gamma), (\zeta), (\mu)$\\ 
  type 3&$\ell_3=(u: av: bv: c v^2: \eta^5 u^5)$ &(i5)(kl4) & $(\beta), (\epsilon), (\nu)$\\
  type 4&$\ell_4=(av: bv: cv: u^2: \eta^5 u^5)$ &(45)(klm) & $(\rho)$\\ 
  \hline
\end{tabular}\\[8pt]
subject to the conditions $\eta^{10}=-1$ and 
\eqn{
\label{CY3ms}
	a^{10}+b^{5}+c^{2}&=0 \quad  \mathbb{P}_{[1,2,5]}\qquad \text{for type 1}\\
	a^{10}+b^{10}+c^{2}&=0 \quad  \mathbb{P}_{[1,1,5]}\qquad \text{for type 2}\\
	a^{10}+b^{10}+c^{5}&=0 \quad  \mathbb{P}_{[1,1,2]}\qquad \text{for type 3}\\
	a^{10}+b^{10}+c^{10}&=0 \quad  \mathbb{P}^2 \qquad\qquad\!\! \text{for type 4}\\
}
The representatives of the equivalence classes of matrix factorisations used are listed in appendix (\ref{App3Fac}). Alone the matrix factorisation $(45)(klm)$ requires some brief comments.
The first factor of this factorisation $(45)$ stands for the formal factorisation $x_4^5+x_5^2=(x_4^5+x_5^2)\cdot 1$. Since this factorisation has an empty cohomology it must be identified with the vacuum configuration \cite{Brunner:2003dc}. This does not mean that the cohomology of the full factorisation $(45)(klm)$ is empty. Rather we should construct factorisations from the reduced superpotential $W-x_4^5-x_5^2$. Effectively this splits off the coordinates $x_4$ and $x_5$ from the boundary sector of the model while not affecting the other coordinates.

The list of possible permutation points is given by
\eqn{
	\text{intersection }1-2:\qquad&(ij)(k4)(5) \\
	\text{intersection }1-3:\qquad&(ij)(k5)(4) \\
	\text{intersection }2-3:\qquad&(i4)(k5)(m) 
}

Generally one would expect an intersection point between branches 1 and 4 of the form $(ij)(45)(m)$. It turns out that this matrix factorisation is not directly accessible as an intersection of two branches. To see this we derive from the parametric line equation $\ell_1$ the vanishing polynomials $J_1=x_1-\eta x_2$, $J_2=b^2 x_3^2-a x_4$, $J_3=c^{2}x_4^5-b^{5}x_5^2$. This factorisation has a limit $a\to 0$ which results in $J_1=x_1-\eta x_2$, $J_2=x_3^2$, $J_3=x_4^5+x_5^2$.
The appearance of a quadratic term in the polynomial $J_2$ is interesting, because this point also lies on the branch 4, but the factorisations derived from $\ell_4$ are linear in $x_3$. Therefore there is no connection between the matrix factorisation associated to $(\rho)$ and any of the other branches.

At the permutation points we find the following spectrum of marginal states:
\eqn{
\label{CY3specperm}
	\text{permutation point 1-2:}\quad
	f_1 &= [0]_b \,\odot [1]_b \,\odot [0]_f = x_k^5 (\pi^3-\bar\pi^3) \\
	f_2 &= [\tfrac{1}{5}]_b \odot [\tfrac{4}{5}]_b \odot [0]_f = x_ix_k^4 (\pi^3-\bar\pi^3) \\
	f_3 &= [\tfrac{2}{5}]_b \odot [\tfrac{3}{5}]_b \odot [0]_f = x_i^2x_k^3 (\pi^3-\bar\pi^3) \\
	f_4 &= [\tfrac{3}{5}]_b \odot [\tfrac{2}{5}]_b \odot [0]_f = x_i^3x_k^2 (\pi^3-\bar\pi^3) \\
	f_5 &= [\tfrac{4}{5}]_b \odot [\tfrac{1}{5}]_b \odot [0]_f = x_i^4x_k (\pi^3-\bar\pi^3) \\
	f_6 &= [1]_b \,\odot [0]_b\, \odot [0]_f = x_i^5 (\pi^3-\bar\pi^3) \\
	\text{permutation point 1-3:}\quad
	f_1 &= [0]_b \,\odot [\tfrac{3}{5}]_b \odot [\tfrac{2}{5}]_f = x_k^2 (\pi^3-x_4^2\bar\pi^3) \\
	f_2 &= [\tfrac{1}{5}]_b \odot [\tfrac{1}{5}]_b \odot [\tfrac{2}{5}]_f = x_ix_k (\pi^3-x_4^2\bar\pi^3) \\
	f_3 &= [\tfrac{3}{5}]_b \odot [0]_b \,\odot [\tfrac{2}{5}]_f = x_i^2 (\pi^3-x_4^2\bar\pi^3) \\
	\text{permutation point 2-3:}\quad
	f_1 &= [\tfrac{1}{5}]_b \,\odot [0]_b \odot [\tfrac{4}{5}]_f = x_i (\pi^3-x_m^8\bar\pi^3) \\
	f_2 &= [0]_b \,\odot [\tfrac{1}{5}]_b \odot [\tfrac{4}{5}]_f = x_k (\pi^3-x_m^8\bar\pi^3) \\
}

\subsection{Lines and the moduli web}

On type 1 branches $(ij)(4l5)$ the unobstructed fermion $\partial_b Q$ is proportional to $x_l^2$. This allows us to construct two additional fermions so that the marginal cohomology is given by
\eqn{
	\psi &= \partial_b Q\\
	\psi^\perp &=\frac{x_i}{x_l} \partial_b Q\\
	\psi^{\perp\!\!\!\perp} &=\frac{x_i^2}{x_l^2} \partial_b Q\ .
}
This is indeed the full cohomology because from (\ref{CY3specperm}) we see that branches of type 1 intersect branches of type 3 with a three-dimensional cohomology.

On the other branches we find only two fermions, which are given by
\eqn{
	\psi &= \partial_b Q\\
	\psi^\perp &=\frac{x_i}{x_l} \partial_b Q
}
for both, branches of second and third type. Again, (\ref{CY3specperm}) tells us that this is the full cohomology since the spectrum at the permutation point 1-3 is two-dimensional.

The joining relations at the points 1-2 between the branches $Q=(ij)(5l4)$ and $Q'=(l4)(5ij)$ are
\eqna{
	&\psi\;\;\; \longrightarrow &f_1 \nonumber\\
	&\psi^\perp\, \longrightarrow &f_2 \nonumber\\
	&\psi^{\perp\!\!\!\perp}\longrightarrow &f_3\nonumber\\
	&&f_4\\
	&&f_5 \longleftarrow {\psi'}^\perp \nonumber\\
	&&f_6 \longleftarrow {\psi'} \nonumber
}
For the points 1-3 between $Q=(ij)(4l5)$ and $Q'=(l5)(4ij)$ we find
\eqna{
	&\psi\;\;\; \longrightarrow &f_1 \nonumber\\
	&\psi^\perp\, \longrightarrow &f_2 \longleftarrow {\psi'}^\perp \\
	&\psi^{\perp\!\!\!\perp}\longrightarrow &f_3 \longleftarrow {\psi'} \ .\nonumber
}
Finally, for the points 2-3 between $Q=(k4)(il5)$ and $Q'=(l5)(ik4)$ the joining relations are
\eqna{
	&\psi\;\;\; \longrightarrow &f_1\longleftarrow {\psi'}^\perp \nonumber\\
	&\psi^\perp\, \longrightarrow &f_2\longleftarrow {\psi'} \ .
}

All states at the permutation points can be continued on the branches, except $f_4$. We briefly discuss their obstructions.

At the intersection point 1-2, $(ij)(k4)(5)$, the first order condition for deformations in directions $f_3$, $f_4$ and $f_5$ are obstructed because
\eqn{
	-\frac{1}{2}\{f_3,f_3\} &= x_i^4x_k^6 \qquad
	-\frac{1}{2}\{f_4,f_4\} = x_i^6x_k^4 \qquad
	-\frac{1}{2}\{f_5,f_5\} = x_i^8x_k^2 \ ,
}
which are all non-trivial in the cohomology. Thus, no further correction to the matrix factorisation can be found, and the direction is obstructed.

For $f_2$ the obstruction does not occur until the forth order. The second order correction $Q_2$ that solves
\eqn{
	\{Q_0, Q_2\} &= -\frac{1}{2}\{ f_2, f_2\} = x_i^2x_k^8
}
is given by
\eqn{
	Q_2 &=\frac{\eta'}{5}x_i^2\left(\pi^2-\sum_{i=0}^3{\eta'}^{2i}(i+1)x_3^{2i}x_4^{3-i}\right)\ .
}
%
%
For the third order correction we then have the equation
\eqn{
        \{Q_0, Q_3\} &= -\{f_2,Q_2\}=0, \
}
which is solved by $Q_3=0$. At the fourth order we encounter the obstructed equation
\eqn{
        \{Q_0, Q_4\} &= -\{f_2,Q_3\}-\frac{1}{2}\{Q_2,Q_2\}\nonumber\\
                     &=-\frac{{\eta'}^{6}x_i^4}{50}\sum_{i=0}^3{\eta'}^{2i}(i+1)x_k^{2i}x_4^{3-i}\ .
}
The right hand side of the above equation is gauge equivalent to \(\lambda x_i^4x_k^6\), \(\lambda\in\mathbb{C}\),
which is a non-trivial element of the cohomology. The fermion \(f_2\) is therefore obstructed at fourth order.

\subsection{Bulk perturbations and effective superpotentials}

The bulk deformations which switch on the various fermions are listed below:
\eqn{
	\text{type 1 branch }(ij)(4l5):\qquad\qquad\qquad&\\
	\vev{s^{(8)}(x_i, x_j)s^{(2)}(x_l, x_4, x_5) \; \psi} &= \frac{\eta s^{(8)}(\eta,1)}{10} \frac{b s^{(2)}(1,b,c)}{c}\\
	\vev{s^{(7)}(x_i, x_j)s^{(3)}(x_l, x_4, x_5) \; \psi^\perp} &= \frac{\eta^2 s^{(7)}(\eta,1)}{10} \frac{b s^{(3)}(1,b,c)}{c}\\
	\vev{s^{(6)}(x_i, x_j)s^{(4)}(x_l, x_4, x_5) \; \psi^{\perp\!\!\!\perp}} &= \frac{\eta^3 s^{(6)}(\eta,1)}{10} \frac{b s^{(4)}(1,b,c)}{c}
}
\eqn{
	\text{type 2 branch }(i4)(kl5):\qquad\qquad\qquad&\\
	\vev{s^{(7)}(x_i, x_4)s^{(3)}(x_l, x_k, x_5) \; \psi} &= -\frac{\eta s^{(7)}(\eta,1)}{10} \frac{ s^{(3)}(1,b,c)}{c}\\
	\vev{s^{(6)}(x_i, x_4)s^{(4)}(x_l, x_k, x_5) \; \psi^\perp} &= -\frac{\eta^2 s^{(6)}(\eta,1)}{10} \frac{s^{(4)}(1,b,c)}{c}
}
\eqn{
	\text{type 3 branch }(i5)(kl4):\qquad\qquad\qquad&\\
	\vev{s^{(4)}(x_i, x_5)s^{(6)}(x_l, x_k, x_4) \; \psi} &= \frac{\eta s^{(4)}(\eta,1)}{10} \frac{ s^{(6)}(1,b,c)}{c^4}\\
	\vev{s^{(3)}(x_i, x_5)s^{(7)}(x_l, x_k, x_4) \; \psi^\perp} &= \frac{\eta^2 s^{(3)}(\eta,1)}{10} \frac{s^{(7)}(1,b,c)}{c^4}\\
}
From this we can immediately derive the points in the moduli space for which branes deform with bulk moduli by requiring that the r.h.s.\ of the $\psi$-correlators $\vev{G \psi}$ vanish. There are 20 such points on type 1 branches, 30 on type 2 and 60 on type 3, at which the matrix factorisations can be deformed with a bulk modulus.
	
In complete agreement with the previously studied Calabi-Yaus we find that the holomorphic functions on the moduli spaces
(\ref{CY3ms}) are a basis for bulk-boundary correlators $\vev{G \psi}$:

The curve $a^{10}+b^5+c^2=0$ (type 1) has genus $g=2$ and we find the holomorphic functions \cite{Griffiths}
\eqn{
	\left\{\frac{b^sc^s}{c}, 0\le 2s+5c\le 2 \right\}\ .
}
For the curve $a^{10}+b^{10}+c^2=0$ (type 2) the genus is $g=4$. The holomorphic functions are
\eqn{
	\left\{\frac{b^sc^s}{c}, 0\le s+5c\le 3 \right\}\ .
}
Finally the curve  $a^{10}+b^{10}+c^5=0$ (type 3) has genus $g=16$ and its holomorphic functions are
\eqn{
	\left\{\frac{b^sc^s}{c}, 0\le s+2c\le 6 \right\}\ .
}

Integration of the bulk-boundary correlators leads to the effective superpotentials
\eqn{
	{\cal W}_{eff}^\text{type 1} &= \lambda  \sum_{i+2j+5k=2} s^{(2)}_{ijk} {\cal W}_{2(j+1),5} \\
	{\cal W}_{eff}^\text{type 2} &= \lambda  \sum_{i+2j+5k=2} s^{(3)}_{ijk} {\cal W}_{j+1,5} \\
	{\cal W}_{eff}^\text{type 3} &= \lambda  \sum_{i+2j+5k=2} s^{(6)}_{ijk} {\cal W}_{j+1,2(k+1)} \ ,
}
where
\eqn{
	{\cal W}_{rs}=\frac{b^r}{r}{}_2{\sf{F}}_1\left(\frac{r}{10},1-\frac{s}{10};1+\frac{r}{10};-{b}^{10}\right)\ .
}

\section{Conclusions}

In this article we have extended the work of \cite{Baumgartl:2007an} to the set of one-parameter Calabi-Yaus.
It has been shown that the moduli space of lines in these manifolds consists of several branches which are connected at permutation points. These points are distinguished by their enhanced spectrum of marginal states, coming once from the two different fermions generating the flat directions and also from fermions that are marginal only at the permutation points. The underlying symmetry group which is given by the soccer ball diagram is universal for all models that are tensor products of five minimal models; we have seen that this symmetry is modified in weighted space. The joining conditions which determine how the marginal spectra on the branches are connected at permutation points are non-trivial and have been explicitly computed. It is important to understand these conditions in order to get a global view of the moduli space.

It is very interesting to study the lifting of open string moduli under closed string deformations. 
Under bulk deformations we found that each branch of the moduli space collapses into a set of discrete points which are extrema of an effective potential. There the branes are stable and deform along with complex structure deformations to finite coupling. Since the boundary moduli are now fixed by the bulk moduli the large complex structure limit is accessible. This in principle makes it possible to apply methods as presented in \cite{Walcher:2006rs} in order to find more examples for open-closed Picard-Fuchs equation. This could in particular be interesting since on each branch a whole set of branes deforms with $W$, so that the domain wall tensions between various branes can be computed.

For unfixed boundary modulus we have computed explicitly the bulk-induced effective potential for the holomorphic sector of the B-model by integrating the bulk-boundary correlators. It is an essential point to see that this method works not only in the simplest case for the Fermat quintic, but also for more complicated models. In particular we have shown that the bulk-boundary correlators are in one-to-one correspondence to holomorphic differentials on the Riemann curve forming the moduli space. Hence a very concrete geometrical interpretation is assigned to them. 

This correspondence is very interesting because it seems that topological data on the moduli space, namely the genus, can be extracted from a computation of bulk-boundary correlators.
In practice the genus can correctly be obtained only at generic points, where no bulk-boundary coefficients accidentally vanish. Also, knowledge of the exactly marginal fermion is necessary, which is in a way a `global' information that enters here. In practice it might often be obvious which of the fermions on the D5-moduli space are obstructed, so that the exactly marginal one can be identified without knowledge of the full deformation theory.

Our calculations are conducted at first order in the bulk moduli. For bulk deformations of higher order we expect that the holomorphic differentials acquire corrections, which should give a hint on a modified moduli space, maybe in a similar way as the first order bulk-boundary correlators determine the genus of the open string moduli space. It seems desirable but unreasonable to attempt to take these computations to higher order, since in our approach we have to keep all possible bulk moduli. For practical purposes this is far too complicated; in order to make progress in this direction the number of bulk moduli could be reduced by dividing out some symmetry. However, the most simple situation in which the Fermat quintic is divided by the diagonal symmetry, is exactly the case where there are no boundary moduli generated by $\psi$. Rather, in this situation $\psi^\perp$ is the important fermion. Since $\psi^\perp$ is not exactly marginal we expect a complicated combined bulk-boundary moduli space.

\paragraph{Acknowledgements}

We would like to thank Ilka Brunner, Matthias Gaberdiel and Peter Mayr. The work of M.~B.~is 
supported by an EURYI award and that of S.~W.~by the SNF. Some of the
calculations presented here are based on the masters thesis of S.~W.

\newpage
\appendix
\section{Appendix}

\subsection{Factorisations on {$\mathbb P_{(1,1,1,1,1)}[5]$}}
\label{App0Fac}
On the Fermat quintic the matrix factorisation can be generally written as
\eqna{
	(ij)(klm):
	&J_1=x_i - \eta x_j\qquad\qquad&E_1=\phantom{-}\prod_{\eta'\neq \eta}(x_i-\eta'x_j) \nonumber\\
	&J_2=a x_k-b x_l	 \qquad\qquad& E_2=\phantom{-}\sum_{i=0}^4\frac{b^{4-i}}{a^{5-i}}x_l^{4-i}x_k^i \nonumber\\
	&J_3=c x_l-a x_m \qquad\qquad& E_3=-\sum_{i=0}^4\frac{c^{i}}{a^{i+1}}x_m^{4-i}x_l^i\nonumber\\
}

\subsection{Factorisations on {$\mathbb P_{(1,1,1,1,2)}[6]$}}
\label{App1Fac}
On $\mathbb P_{(1,1,1,1,2)}[6]$ we consider the matrix factorisations
\eqna{
	(5j)(klm):
	&J_1=x_5 - \eta^2 x_j^2\qquad\qquad&E_1= \phantom{-}\prod_{{\eta'}^2\neq \eta^2}(x_5-\eta^2x_j^2) \nonumber\\
	&J_2=a x_k-b x_l	 \qquad\qquad& E_2=\phantom{-}\sum_{i=0}^5\frac{b^{5-i}}{a^{6-i}}x_l^{5-i}x_k^i \nonumber\\
	&J_3=c x_l-a x_m \qquad\qquad& E_3=-\sum_{i=0}^5\frac{c^{i}}{a^{i+1}}x_m^{5-i}x_l^i \nonumber\\
	\nonumber\\
	(ij)(kl5):
	&J_1=x_i - \eta x_j\qquad\qquad&E_1= \phantom{-}\prod_{\eta'\neq \eta}(x_i-\eta'x_j) \nonumber\\
	&J_2=a x_k-b x_l\qquad\qquad&E_2=\phantom{-}\sum_{i=0}^5\frac{b^{5-i}}{a^{6-i}}x_l^{5-i}x_k^i\nonumber\\
  	&J_3=c x_l^2-a^2 x_5\qquad\qquad&E_3=-\sum_{i=0}^2\frac{c^{2-i}}{a^{6-2i}}x_l^{4-2i}x_5^i \nonumber\\
}

\subsection{Marginal spectrum on {$\mathbb P_{(1,1,1,1,2)}[6]$}}
\label{CY1branches}
On branches of type 1 the matrix factorisation takes the form $Q=Q_1\odot Q_2 \odot Q_3$. The first factor $Q_1$ is defined by the polynomial $J_1=x_i-\eta x_j$, so that this part is independent of the boundary moduli. The spectrum of $Q_1$ consists of three bosons with charges 0, $\frac{1}{3} \dots \frac{4}{3}$. In order to construct factorisations of total charge one we look for fermions with charges 1, $\frac{2}{3}$ and $\frac{1}{3}$ in the reduced factorisation $Q'=Q_2 \odot Q_3$.
\subsubsection{Fermions of charge 1}
The ansatz for a general fermion is\footnote{
One might wonder if higher powers of $\pi$ and $\bar\pi$ can appear. In general, these must indeed be taken into account. But here all such higher order terms are trivial because the matrix factorisation itself contains only linear terms. For instance, for a fermion with a term 
$p_{ijk}\pi^i\pi^j\bar\pi^k+q_{ijk}\pi^i\bar\pi^j\bar\pi^k$ its closedness condition becomes 
$\pi^i\pi^j J^k p_{ijk} + \bar\pi^i\bar\pi^j E^k q_{kij} + \pi^i\bar\pi^j(2J^k q_{ijk} + 2E^k p_{kij})$. 
The first term leads to the condition $p_{232}J^2 + p_{233}J^3=0$. Since $J^2$ and $J^3$ are linear independent,
$p_{ijk}$ must be zero. The same argument sets $q_{ijk}$ to zero.
}
\eqn{
\label{appferm}
	\Psi&=p_2\pi^2+m_2\bar\pi^2+p_3\pi^3+m_3\bar\pi^3\ .
}
In this expression $p_i$ and $m_i$ are polynomials in the variables $x_3$, $x_4$, $x_5$. We list their charges and the corresponding number of free parameters (taking into account the higher charge of $x_5$):
\eqna{
	{[}p_2] &= \tfrac{1}{3}\qquad\qquad&\text{parameters: }2\nonumber\\
	{[}p_3] &= \tfrac{4}{3}\qquad\qquad&\text{parameters: }9\nonumber\\
	{[}m_2] &= \tfrac{5}{3}\qquad\qquad&\text{parameters: }12\nonumber\\
	{[}m_3] &= \tfrac{2}{3}\qquad\qquad&\text{parameters: }4\ .
}
Thus the space of such fermions has dimension 27.
The closedness equations become
\eqn{
	p_2E_2+p_3E_3+m_2J_2+m_3J_3=0\ .
}
For general values of $a$, $b$ and $c$ this equation supplies 16 constraints\footnote{For special values of  $a$, $b$ and $c$ some equations might become dependent. This might lead to less constraints and a higher dimension of the kernel for a finite set of points.}. Therefore
\eqn{
	\text{dim} \left( \text{Ker} (Q_2\odot Q_3)\right) = 11\ .
}
In order to determine the dimension of the exact fermions we make the ansatz
\eqn{
	\Lambda &= \lambda_1 + \lambda_2\pi^2\bar\pi^2 + \lambda_3\pi^2\pi^3 +
	\lambda_4\pi^2\bar\pi^3 + \lambda_5\bar\pi^2\pi^3\\ 
	&+ \lambda_6\bar\pi^2\bar\pi^3 
	+\lambda_7\pi^3\bar\pi^3 + \lambda_8\pi^2\bar\pi^2 \pi^3\bar\pi^3\ .
}
The parameters $\lambda_1$ and $\lambda_8$ can't be used to build fermions (\ref{appferm}) hence we can set the to zero. The charge of $\Lambda$ must be zero, thus we get the following list of charges and free parameters:
\eqna{
	{[}\lambda_2] &= 0 \qquad\qquad&\text{parameters: }1\nonumber\\
	{[}\lambda_5] &= 1 \qquad\qquad&\text{parameters: }6\nonumber\\
	{[}\lambda_6] &= \frac{1}{3} \qquad\qquad&\text{parameters: }2\nonumber\\
	{[}\lambda_7] &= 0 \qquad\qquad&\text{parameters: }1\ .
}
There are no $\lambda_3$ and $\lambda_4$ that could meet the charge constraints, so they have been set to zero.
In total we get
\eqn{
	\text{dim}(\text{Im}(Q_2\odot Q_3)) &= 10\ .
}
Thus the cohomology has dimension
\eqn{
	h (Q_2\odot Q_3)= 1\ .
} 
Thus there is generically one such fermion in the spectrum.

\subsubsection{Fermions of charge $\frac{2}{3}$}

The above computations are repeated for fermions of charge $\tfrac{2}{3}$. In this case we  get the following charge tables:
\eqna{
	{[}p_2] &= 0 \qquad\qquad&\text{parameters: }1\nonumber\\
	{[}p_3] &= 1\qquad\qquad&\text{parameters: }6\nonumber\\
	{[}m_2] &= \tfrac{4}{3}\qquad\qquad&\text{parameters: }9\nonumber\\
	{[}m_3] &= \tfrac{1}{3} \qquad\qquad&\text{parameters: }2\ .
}
The closedness condition imposes 12 constraints, thus
\eqn{
	\text{dim} \left( \text{Ker} (Q_2\odot Q_3)\right) = 6\ .
}
For $\Lambda$ with charge $-\tfrac{1}{3}$ we find
\eqna{
	{[}\lambda_5] &= \tfrac{2}{3} \qquad\qquad&\text{parameters: }4\nonumber\\
	{[}\lambda_6] &= 0 \qquad\qquad&\text{parameters: }1
}
with all other $\lambda_i=0$.
From this
\eqn{
	\text{dim}(\text{Im}(Q_2\odot Q_3)) &= 5
}
and 
\eqn{
	h(Q_2\odot Q_3) = 1\ .
}

\subsubsection{Fermions of charge $\frac{1}{3}$}

Now we focus on the fermions of charge $\tfrac{1}{3}$. The charge tables are:
\eqna{
	{[}p_3] &= \tfrac{2}{3}\qquad\qquad&\text{parameters: }4\nonumber\\
	{[}m_2] &= 1\qquad\qquad&\text{parameters: }6\nonumber\\
	{[}m_3] &= 0 \qquad\qquad&\text{parameters: }1
}
with $p_2=0$.
The closedness condition imposes 9 constraints, thus the kernel has dimension zero and the cohomology is empty.

\subsubsection{Fermions of charge 0}

For fermions with zero charge we get
\eqna{
	{[}p_3] &= \tfrac{1}{3}\qquad\qquad&\text{parameters: }2\nonumber\\
	{[}m_2] &= \tfrac{2}{3}\qquad\qquad&\text{parameters: }4
}
with $p_2=0=m_3$.
The closedness condition supplies 6 constraints, thus again the kernel has dimension zero and the cohomology is empty.

In total we find that
\eqn{
	h (Q) = 2\ ,
}
thus there are two fermions defined on the type 1 branches. Since we have found two fermions $\psi=\partial_bQ$ and $\psi^\perp=\frac{x_i}{x_k}\partial_bQ$ we have found the maximum number and thus the complete marginal cohomology.

\subsection{Obstructions on type-1-branches of {$\mathbb P_{(1,1,1,1,2)}[6]$}}
\label{CY1obstr}

Since the three-point function
\eqn{
	\vev{(\psi^\perp)^3} = 0
}
everywhere on the branches $(ij)(kl5)$ we expect obstruction to appear at higher order.
A perturbation of $Q$ with $\psi^\perp$ leads to the ansatz
\begin{align}
  Q(\lambda)=\sum_{n}\lambda^n Q_n
\end{align}
with $Q_0=Q$ and $Q_1=\psi^\perp$. We know already that $\psi^\perp$ is in the cohomology, so the first order equation is already satisfied.
The second order equation
\begin{align}
  \{Q_0,Q_2\}&=-\frac{1}{2}\{\psi^\perp,\psi^\perp\}\nonumber\\
  	&=\frac{1}{4\hat\eta^{10}}\left(\sum_{i=0}^4(5-i)b^{4-i}c^8x_1^2x_3^{4-i}x_4^i \right.\nonumber\\
  	&\quad\quad\;\;\;\left.+2\sum_{i=0}^1(4-2i)b^{10}c^{2-2i}x_1^2x_3^{4-2i}x_5^i\right)
\end{align}
is solved by
\eqn{
	Q_2 &= -\left(5\frac{b^{10}}{c^5}+5\frac{b^4}{c^2}\right) x_i^2\pi^3\\
		&+\sum_{n=0}^3\frac{(n+1)(n+2)}{2}b^{3-n}x_i^2x_l^{3-n}x_k^n \bar\pi^2\\
		&+\left(\left(4\frac{b^{10}}{c^5}+5\frac{b^4}{c^2}\right)x_i^2x_5
			+ \left(4\frac{b^{10}}{c^4}+10\frac{b^4}{c}\right)x_i^2x_l^2\right)\bar\pi^3\ .
}
With the help of computer algebra software we verify that the third order equation
\eqn{
	\{Q_0,Q_3\} &= -\{\psi^\perp,Q_2\}
}
has no solutions.
Therefore the fermion $\psi^\perp$ is obstructed.

\subsection{Factorisations on {$\mathbb P_{(1,1,1,1,4)}[8]$}}
\label{App2Fac}
On $\mathbb P_{(1,1,1,1,4)}[8]$ we consider the matrix factorisations
\eqna{
	(5j)(klm):
	&J_1=x_5 - \eta^4 x_j^4\qquad\qquad&E_1= \phantom{-}\prod_{{\eta'}^4\neq \eta^4}(x_5-\eta^4x_j^4) \nonumber\\
	&J_2=a x_k-b x_l	 \qquad\qquad& E_2=\phantom{-}\sum_{i=0}^7\frac{b^{7-i}}{a^{8-i}}x_l^{7-i}x_k^i \nonumber\\
	&J_3=c x_l-a x_m \qquad\qquad& E_3=-\sum_{i=0}^7\frac{c^{i}}{a^{i+1}}x_m^{7-i}x_l^i \nonumber\\
	\nonumber\\
	(ij)(kl5):
	&J_1=x_i - \eta x_j\qquad\qquad&E_1= \phantom{-}\prod_{\eta'\neq \eta}(x_i-\eta'x_j) \nonumber\\
	&J_2=a x_k-b x_l\qquad\qquad&E_2=\phantom{-}\sum_{i=0}^7\frac{b^{7-i}}{a^{8-i}}x_l^{7-i}x_k^i\nonumber\\
  	&J_3=c x_l^4-a^4 x_5\qquad\qquad&E_3=-\sum_{i=0}^1\frac{c^{1-i}}{a^{8-4i}}x_l^{4-4i}x_5^i \nonumber\\
}

\subsection{Factorisations on {$\mathbb P_{(1,1,1,2,5)}[10]$}}
\label{App3Fac}
On $\mathbb P_{(1,1,1,2,5)}[10]$ we consider the matrix factorisations
\eqna{
	(4j), j\ne 5:
	&J_1=x_4 - \eta^2 x_j^2\qquad\qquad&E_1= \phantom{-}\prod_{{\eta'}^2\neq \eta^2}(x_4-{\eta'}^2x_j^2) \nonumber\\
	\nonumber\\
	(5j), j\ne 4:
	&J_1=x_5 - \eta^5 x_j^5\qquad\qquad&E_1= \phantom{-}\prod_{{\eta'}^5\neq \eta^5}(x_5-{\eta'}^5x_j^5) \nonumber\\
	(kl5), k,l\ne 4:
	&J_2=a x_k-b x_l\qquad\qquad&E_2=\phantom{-}\sum_{i=0}^9\frac{b^{i}}{a^{i+1}}x_k^{9-i}x_l^i\nonumber\\
  	&J_3=c x_l^5-a^5 x_5\qquad\qquad&E_3=-\sum_{i=0}^1\frac{c^{1-i}}{a^{10-5i}}x_l^{5-5i}x_5^i \nonumber\\
	\nonumber\\
	(kl4), k,l\ne 5:
	&J_2=a x_k-b x_l\qquad\qquad&E_2=\phantom{-}\sum_{i=0}^9\frac{b^{i}}{a^{i+1}}x_k^{9-i}x_l^i\nonumber\\
  	&J_3=c x_l^2-a^2 x_4\qquad\qquad&E_3=-\sum_{i=0}^4\frac{c^{4-i}}{a^{10-2i}}x_l^{8-2i}x_4^i \nonumber\\
	\nonumber\\
	(4l5), l\ne 4,5: 
	&J_2=a^2x_4-bx_l^2\qquad\qquad&E_2=\phantom{-}\sum_{i=0}^4\frac{b^{4-i}}{a^{10-2i}}x_l^{8-2i}x_4^i\nonumber\\
	&J_3=cx_l^5-a^5x_5\qquad\qquad&E_3=-\sum_{i=0}^1\frac{c^{1-i}}{a^{10-5i}}x_l^{5-5i}x_5^i\nonumber\\
}


\end{document}